\documentclass[aps,twocolumn,10pt,superscriptaddress,amsfonts,amssymb,amsmath,preprintnumbers,floatfix,showpacs,noeprint]{revtex4-1}
\usepackage{graphicx}
\usepackage{bm}
\usepackage{color,ulem}
\usepackage[dvipdfmx,colorlinks=true,citecolor=magenta,urlcolor=blue]{hyperref}
\begin{document}
\title{Post-growth annealing effects on charge and spin excitations in Nd$_{2-x}$Ce$_x$CuO$_4$}
\author{Kenji Ishii}
\affiliation{Synchrotron Radiation Research Center,
National Institutes for Quantum and Radiological Science and Technology, Hyogo 679-5148, Japan}
\author{Shun Asano}
\affiliation{Institute for Materials Research, Tohoku University, Sendai 980-8577, Japan}
\author{Masumi Ashida}
\affiliation{Graduate School of Science and Technology, Kwansei Gakuin University, Hyogo 669-1337, Japan}
\author{Masaki Fujita}
\affiliation{Institute for Materials Research, Tohoku University, Sendai 980-8577, Japan}
\author{Biqiong Yu}
\affiliation{School of Physics and Astronomy, University of Minnesota, Minneapolis, Minnesota 55455, USA}
\author{Martin Greven}
\affiliation{School of Physics and Astronomy, University of Minnesota, Minneapolis, Minnesota 55455, USA}
\author{Jun Okamoto}
\affiliation{National Synchrotron Radiation Research Center, Hsinchu 30076, Taiwan}
\author{Di-Jing Huang}
\affiliation{National Synchrotron Radiation Research Center, Hsinchu 30076, Taiwan}
\author{Jun'ichiro Mizuki}
\affiliation{Graduate School of Science and Technology, Kwansei Gakuin University, Hyogo 669-1337, Japan}
\date{\today}

\begin{abstract}
We report a Cu $K$- and $L_3$-edge resonant inelastic x-ray scattering study of charge and spin excitations of bulk Nd$_{2-x}$Ce$_x$CuO$_4$,  with focus on post-growth annealing effects.
For the parent compound Nd$_2$CuO$_4$ ($x=0$), a clear charge-transfer gap is observed in the as-grown state, whereas the charge excitation spectra indicate that electrons are doped in the annealed state.
This is consistent with the observation that annealed thin-film and polycrystalline samples of $RE_2$CuO$_4$ (RE = rare earth) can become metallic and superconducting at sufficiently high electron concentrations without Ce doping.
For $x$ = 0.16, a Ce concentration for which it is known that oxygen reduction destroys long-range antiferromagnetic order and induces superconductivity, we find that the high-energy spin excitations of non-superconducting as-grown and superconducting annealed crystals are nearly identical.
This finding is in stark contrast to the significant changes in the low-energy spin excitations previously observed via neutron scattering.
\end{abstract}

\preprint{preprint \today}

\maketitle

\section{Introduction}
High-transition-temperature (high-$T_c$) superconductivity in the cuprates occurs when either holes or electrons are doped into parent antiferromagnetic (AFM) Mott insulators.
Most electron-doped cuprate superconductors have the chemical formula $RE_{2-x}$Ce$_x$CuO$_4$ ($RE$ = La, Pr, Nd, Sm, Eu), with fourfold-coordinated Cu atoms in the so-called $T'$ structure.
Electrons are doped into the quintessential CuO$_2$ planes via partial substitution of trivalent $RE$ with tetravalent Ce.
As-grown crystals do not superconduct, and a post-growth anneal step is necessary for superconductivity to occur, even at high Ce concentrations \cite{Tokura1989,Takagi1989a}.
Because the anneal is performed under reduction conditions, oxygen is removed from the sample and, consequently, additional electrons are doped into the CuO$_2$ planes.
However, as recognized early on \cite{Matsuda1992}, the post-growth annealing step also causes changes in the disorder potential experienced by the itinerant carriers. 
One scenario, which has been confirmed in diffraction experiments \cite{Radaelli1994,Schultz1996}, is the removal of a small density of oxygen atoms from the nominally vacant apical sites located just above/below of the Cu atoms \cite{Matsuda1992}.
These apical-oxygen ``impurities'' may cause local lattice distortions and promote the tendency toward charge-order rather than superconductivity  \cite{Onose1999}.
Since then, it has been established that the low-oxygen-fugacity environment during the annealing process required to create a superconducting state 
causes a partial decomposition (at the $\sim 1\%$ level) of $RE_{2-x}$Ce$_x$CuO$_4$ into ($RE$,Ce)$_2$O$_3$ \cite{Mang2004}. A second scenario involves the existence of Cu vacancies in the as-grown state, and for such vacancies to be repaired as a byproduct of the partial decomposition \cite{Kang2007}.

In the conventional phase diagram of the electron-doped superconductors, the Ce concentration ($x$) is equated with the electron density, and the lower superconducting phase boundary of post-growth-annealed samples is located at $x \sim 0.14$ \cite{Armitage2010}. 
For the archetypal compound Nd$_{2-x}$Ce$_x$CuO$_4$, it was established that the appearance of bulk superconductivity coincides with the disappearance of long-range AFM order and with the emergence of short-range instantaneous spin correlations \cite{Motoyama2007}.
However, it was reported that an improved annealing procedure expands the superconducting phase boundary to lower Ce concentrations in single-crystalline Pr$_{2-x}$Ce$_x$CuO$_4$ \cite{Brinkmann1995} and Pr$_{1.3-x}$La$_{0.7}$Ce$_x$CuO$_4$ \cite{Adachi2013}.
In subsequent studies of thin films and polycrystalline samples with the $T'$ structure, superconductivity was observed even without Ce substitution ($x = 0$) when the oxygen concentration was carefully controlled \cite{Tsukada2005,Takamatsu2012,Krockenberger2013}.
These more recent findings triggered a reexamination of the effects of post-growth annealing \cite{Adachi2017,Song2017}.
A $\mu$SR study \cite{Adachi2016} revealed short-ranged magnetic correlations in superconducting polycrystalline La$_{1.8}$Eu$_{0.2}$CuO$_{4+\delta}$ ($x=0$) and single-crystalline Pr$_{1.2}$La$_{0.7}$Ce$_{0.1}$CuO$_{4+\delta}$ ($x=0.10$).
Both samples were prepared under improved anneal conditions and correspond to the non-superconducting phase ($x<0.14$) in the conventional phase diagram.
Moreover, angle-resolved photoemission spectroscopy (ARPES) measurements of superconducting Pr$_{1.2}$La$_{0.7}$Ce$_{0.1}$CuO$_{4+\delta}$ \cite{Horio2016} showed no signature of a pseudogap, in contrast to previous ARPES work on Nd$_{2-x}$Ce$_x$CuO$_4$ \cite{Armitage2001,Matsui2005}, where such signatures were associated with a significant increase of spin correlations on cooling.

The relationship between magnetism and superconductivity is a central issue in the cuprates, as these oxides are doped AFM Mott insulators and spin-fluctuations are a prominent candidate for the pairing mechanism \cite{Scalapino2012}. As noted, there exists clear evidence for a change in the magnetic state in bulk superconducting samples. Similarly, dynamic charge correlations have long been argued to play a pivotal role in shaping the cuprate phase diagram \cite{Castellani1996,Kivelson2003}. However, possible changes in the high-energy spin and charge dynamics of the electron-doped cuprates as a result of the crucial post-growth-annealing step have not yet been explored.   

In this paper, we aim to clarify the effects of annealing on both the high-energy charge and spin excitations of the the original 
$T'$-structured superconductor Nd$_{2-x}$Ce$_x$CuO$_4$ using resonant inelastic X-ray scattering (RIXS).
During the past two decades, RIXS has gained much attention as a probe of electronic excitations of superconducting cuprates \cite{Ament2011,Ishii2013,Dean2015}.
While inelastic neutron scattering (INS) has been used mostly to study spin excitations below $\sim100$ meV, RIXS at the Cu $L_3$-edge can cover excitations up to and beyond the magnetic zone-boundary energy ($\sim300$ meV) of the undoped parent materials.
In addition, RIXS is sensitive to the charge degree of freedom, and can provide momentum-resolved spectra of charge excitations.
Even though a number of RIXS experiments of electron-doped cuprates were performed at both the Cu $K$-edge \cite{Ishii2005b,Ishii2006,Li2008,Ishii2014,Ishii2019} and Cu $L_3$-edge \cite{Ishii2014,Lee2014,SilvaNeto2016,Dellea2017,SilvaNeto2018,Hepting2018,Lin2020}, the effects of post-growth annealing on the charge and spin excitations have not been explored.
In order to minimize systematic errors associated with possible variations in the Ce concentration, we performed RIXS experiments of both as-grown and annealed crystals of Nd$_{2-x}$Ce$_x$CuO$_4$ from the same growth.
Charge and spin excitations were measured at the Cu $K$- and $L_3$-edges, respectively.
The improved post-growth annealing procedure that results in superconductivity at low Ce concentration was previously applied to bulk-crystalline Pr$_{2-x}$Ce$_x$CuO$_4$ \cite{Brinkmann1995} and Pr$_{1.3-x}$La$_{0.7}$Ce$_x$CuO$_4$ \cite{Adachi2013}.
However, these compounds are not suitable for Cu $L_3$-edge RIXS studies, because the Pr $M_5$-edge is very close to the Cu $L_3$-edge.
We therefore selected Nd$_{2-x}$Ce$_x$CuO$_4$ for the present work and chose the conventional anneal protocol.

\section{Experiments}
Single crystals of Nd$_{2-x}$Ce$_x$CuO$_4$ with $x$ = 0, 0.05, and 0.16 were grown by the traveling-solvent floating-zone
technique.
Although the oxygen stoichiometry is not exactly 4, we use this chemical formula for simplicity.
We applied the conventional anneal protocol: for each crystal, an as-grown piece was set aside and another piece was annealed in flowing Ar gas for 10 hours at 800$^{\circ}$C ($x=0$) and 900$^{\circ}$C ($x=0.05$ and 0.16).
We measured the magnetic susceptibility of an annealed $x=0.16$ crystal and determined the onset of the superconducting transition to be 15 K.

\begin{table}
\centering
\caption{Lattice constants of the samples at 10 K.}
\label{tab:lat}
\begin{ruledtabular}
\begin{tabular}{p{0.5\columnwidth}p{0.2\columnwidth}p{0.2\columnwidth}}
sample&a (\AA)&c (\AA)\\
\hline
$x=0$ as-grown&3.9363&12.1382\\
$x=0$ annealed&3.9363&12.1334\\
$x=0.05$ as-grown&3.9360&12.1083\\
$x=0.05$ annealed&3.9365&12.1072\\
$x=0.16$ as-grown&3.9369&12.0422\\
$x=0.16$ annealed&3.9397&12.0473\\
\end{tabular}
\end{ruledtabular}
\end{table}

The Cu $K$-edge RIXS measurement of the charge excitations was carried out at beam line 11XU of SPring-8 in Japan.
Incident x-rays were monochromatized by a Si(111) double-crystal monochromator and a Si(400) channel-cut monochromator.
$\pi$-polarized x-rays with 8991 eV were incident on the $ab$-plane of the samples, and horizontally scattered photons were energy-analyzed by a Ge(733) analyzer.
The total energy resolution was 270 meV (full-width-at-half-maximum).
All spectra were taken at a low temperature of about 10 K.
In order to minimize elastic scattering, we selected the absolute momentum transfer so that the scattering angle ($2\theta$) was close to 90$^{\circ}$.
We calculated the momentum transfer from the lattice constants determined from the (0,0,14) and (-1,0,13) Bragg reflections.
The lattice constants are summarized in Tab.~\ref{tab:lat}.

The Cu $L_3$-edge RIXS measurements of the spin excitations were performed at Taiwan Light Source (TLS) beam line 05A1 of the National Synchrotron Radiation Research Center (NSRRC) using an AGM-AGS spectrometer \cite{Lai2014}.
The incident photon energy was tuned to the peak of the x-ray absorption spectrum and $\pi$-polarized x-rays were used.
The single-crystal samples were cleaved in air just before installing them in the vacuum chamber, and the x-rays were incident on the cleaved $ab$-plane.
The scattering angle was kept at 130$^{\circ}$, and the two-dimensional in-plane momentum transfer ($\bm{q}$) was changed by rotating the crystal about the axis perpendicular to the horizontal scattering plane.
Positive values of $h$ in $\bm{q}=(h,0)$ corresponds to sample rotation toward the grazing exit condition.
Because the intensity of spin excitations is larger for $h>0$ than for $h<0$ due to a well-known polarization effect \cite{Ament2009b,Sala2011}, we mainly obtained spectra at $h>0$.
All the spectra were taken at a low temperature of about 20 K.
We used the lattice constants determined in the Cu $K$-edge experiments for the calculation of momentum transfer.

\section{Results}

\begin{figure*}[t]
\centering
\includegraphics[scale=1.0]{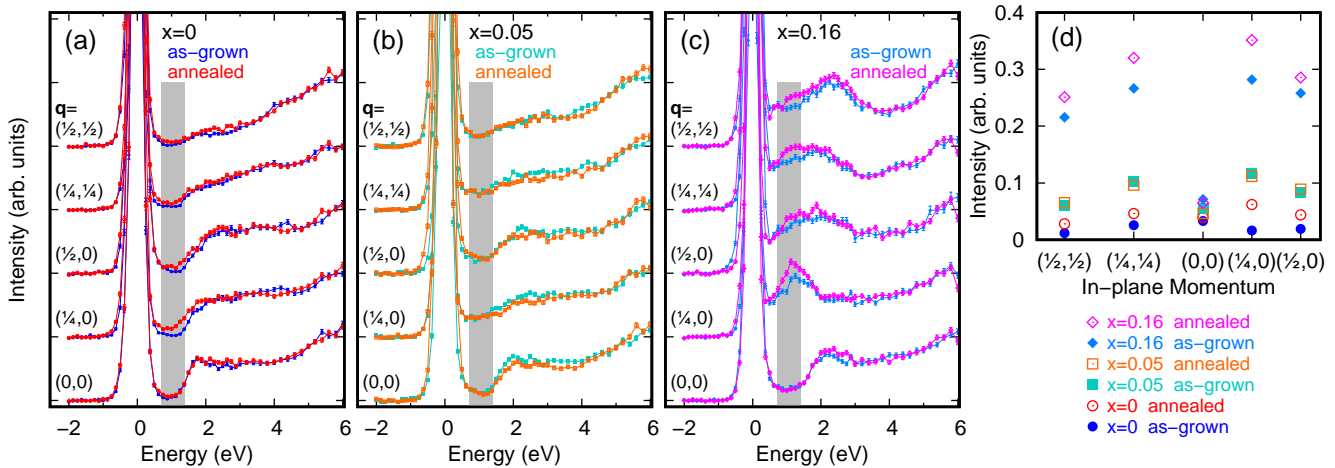}
\caption{(color online). Cu $K$-edge RIXS spectra for Nd$_{2-x}$Ce$_x$CuO$_4$ with (a) $x=0$, (b) $x=0.05$, and (c) $x=0.16$.
Filled and open symbols indicate the spectra of as-grown and annealed crystals, respectively.
Spectra are normalized to the integrated intensity in the 3.5-6.0 eV range (see text).
(d) Integrated intensity of the intraband excitations. The integration range is 0.7-1.4 eV, as indicated by grey bands in (a)-(c).}
\label{fig:cu-k}
\end{figure*}

\subsection{Charge excitations in Cu $K$-edge RIXS}
First, we consider the charge excitations observed with Cu $K$-edge RIXS.
Figures \ref{fig:cu-k}(a), (b) and (c) compare the Cu $K$-edge RIXS spectra for as-grown and annealed crystals with $x=0$, 0.05 and 0.16, respectively.
The momentum transfer along [001] is 12.65 r.l.u.~for $x=0$, 12.54~r.l.u.~for as-grown $x=0.05$, and 12.55 r.l.u.~for annealed $x=0.05$ and 0.16 (1 r.l.u.~$= 2\pi/c$).
The spectra are normalized to the integrated intensity in the 3.5-6.0 eV range, where the spectral weight of high-energy molecular orbital excitations dominates \cite{Hill1998,Kim2004b}.

Before considering the annealing effect in the present data, we briefly mention the previously observed spectral change induced by the substitution of Ce for Nd \cite{Ishii2005b,Ishii2006}.
Charge excitations below $\sim$ 3 eV in Nd$_{2-x}$Ce$_x$CuO$_4$ consist of interband excitations across the charge-transfer gap and intraband excitations of doped electrons.
The former, situated at $\sim 2$ eV, are dominant at $\bm{q}=(0,0)$ and nearly unchanged upon doping.
The latter appear at nonzero in-plane momentum [${\bm q} \ne (0,0)$] and their intensity is roughly proportional to the Ce concentration, and therefore can be regarded as a measure of the doped electron concentration.

Figure~\ref{fig:cu-k}(a) shows for the as-grown $x=0$ crystal an absence of spectral weight around 1 eV at all measured momenta.
This is a clear feature of the charge-transfer gap.
In contrast, nonzero weight emerges below 2 eV in the spectra of the corresponding annealed crystal and is most prominent along [1,0].
Because this spectral change is qualitatively similar to that found upon Ce substitution \cite{Ishii2005b,Ishii2006}, this confirms that electrons are doped in the annealed crystal, even without Ce substitution.

For $x=0.05$, the spectra around 1 eV do not show an appreciable annealing effect on the intraband excitations, as shown in Fig.~\ref{fig:cu-k}(b).
The likely reason for the absence of an effect is that, for this doping level, the chosen annealing condition is insufficient to induce a substantial change in electronic states.
The slight change observed in the 2-3 eV range might either result from annealing effects on the interband excitations across the charge-transfer gap or from an imperfect intensity normalization.
In any case, the result is different from what is observed for $x=0$ and 0.16.
Although the annealing condition is the same as for $x=0.05$, we observe a clear change for $x=0.16$; Fig.~\ref{fig:cu-k}(c) shows that the spectral weight of the intraband excitations of the annealed crystal is larger than that of the as-grown crystal.
At this Ce concentration, the intraband excitations form a broad peak whose high-energy tail extends to 2-3 eV, and the annealing effect is dominant on the low-energy side.

In Fig.~\ref{fig:cu-k}(d), we plot the RIXS intensity integrated in the 0.7-1.4 eV range.
This energy range is selected to avoid the interband excitations and the tail of elastic scattering.
As noted, the intensity for $x=0$ and 0.16 increases for $\bm{q} \ne (0,0)$.
For $x=0$, the annealing effect is 1/3 to 1/2 of the intensity change seen upon 5\% Ce substitution, whereas for $x=0.16$, the change is comparable to that upon 5\% Ce substitution.
Under the assumption that the intensity is proportional to the density of doped electrons, our data indicate that annealing corresponds to a few \% electron doping per Cu atom.
This agrees with the amount of oxygen loss under conventional annealing condition \cite{Uefuji2002}.
Moreover, it is consistent with the previously observed $\sim 3\%$ shift of the magnetic phase boundary in annealed samples \cite{Mang2004a}.

\begin{figure*}[t]
\centering
\includegraphics[scale=1.0]{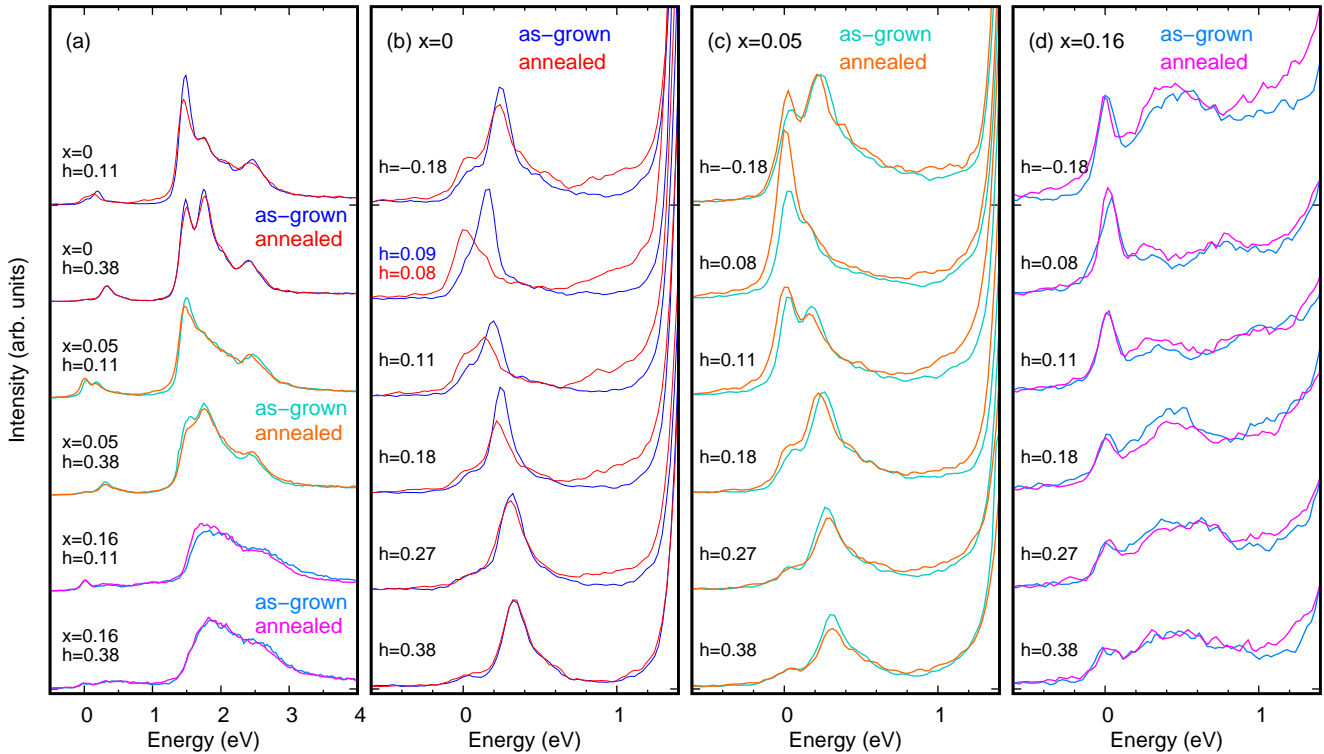}
\caption{(color online). (a)-(d) Cu $L_3$-edge RIXS spectra normalized to the $dd$ intensity in the 1.0-3.5 eV range for as-grown and annealed crystals of Nd$_{2-x}$Ce$_x$CuO$_4$, as indicated.
The intensity of $x=0.16$ in (d) is multiplied by 1.3 to better highlight the relatively broad response.
The in-plane momentum transfer is $\bm{q}=(h,0)$.
Except for $h=0.09$ ($h=0.08$) for the as-grown (annealed) $x=0$ crystal, spectra are compared at the same momentum transfer.}
\label{fig:cu-l3}
\end{figure*}

\subsection{Spin and charge excitations in Cu $L_3$-edge RIXS}
Next, we present the spin and charge excitations observed with Cu $L_3$-edge RIXS.
Figure \ref{fig:cu-l3}(a) shows representative Cu $L_3$-edge RIXS spectra for Nd$_{2-x}$Ce$_x$CuO$_4$ including the energy range of $dd$ excitations.
The spectra are normalized to the integrated intensity of $dd$ excitations in the 1.0-3.5 eV range.
The spectral features of the $dd$ excitations in $x=0$ are typical of the insulating cuprates with fourfold-coordinated Cu$^{2+}$ and three peaks at 1.5, 1.8, and 2.5 eV are ascribed to the transitions $d_{xy} \to d_{x^2-y^2}$, $d_{yz,zx} \to d_{x^2-y^2}$, and $d_{3z^2-r^2} \to d_{x^2-y^2}$, respectively \cite{Sala2011,Kang2019}.
With increasing $x$, the peaks are significantly broadened.
We magnify the energy range of spin and charge excitations in Figs.~\ref{fig:cu-l3}(b)-(d) and compare the spectra for as-grown and annealed crystals.
The main spectral features are consistent with previous studies \cite{Ishii2014,Lee2014}.
The dominant component below 1 eV is due to spin [single (para)magnon] excitations.
In the parent compound ($x=0$), phonon and two-magnon excitations are observed below and above the spin excitations, respectively.
When electrons are doped, the spin excitations are broadened and the two-magnon excitations are hardly identified.
For $x=0.16$, in the low-$q$ region, an additional peak of charge origin appears above the spin excitations.
This feature was previously observed \cite{Ishii2014,Lee2014,Hepting2018} and is connected to the intraband charge excitations in Cu $K$-edge RIXS spectra \cite{Ishii2019}.

Focusing next on the annealing effects, we observe for $x=0$ that the spin excitation peak slightly broadens.
Since the existence of doped electrons in the annealed $x=0$ crystal is indicated in the Cu $K$-edge RIXS data [Fig.~\ref{fig:cu-k}(a)], this broadening is most likely the result of a shortened magnon lifetime due to interactions with the itinerant electrons.
For $x=0.05$ and 0.16, the width is already broad in the as-grown crystals, and it is difficult to discuss the annealing effect on the peak width from the present spectra.

\begin{figure*}
\centering
\includegraphics[scale=1.0]{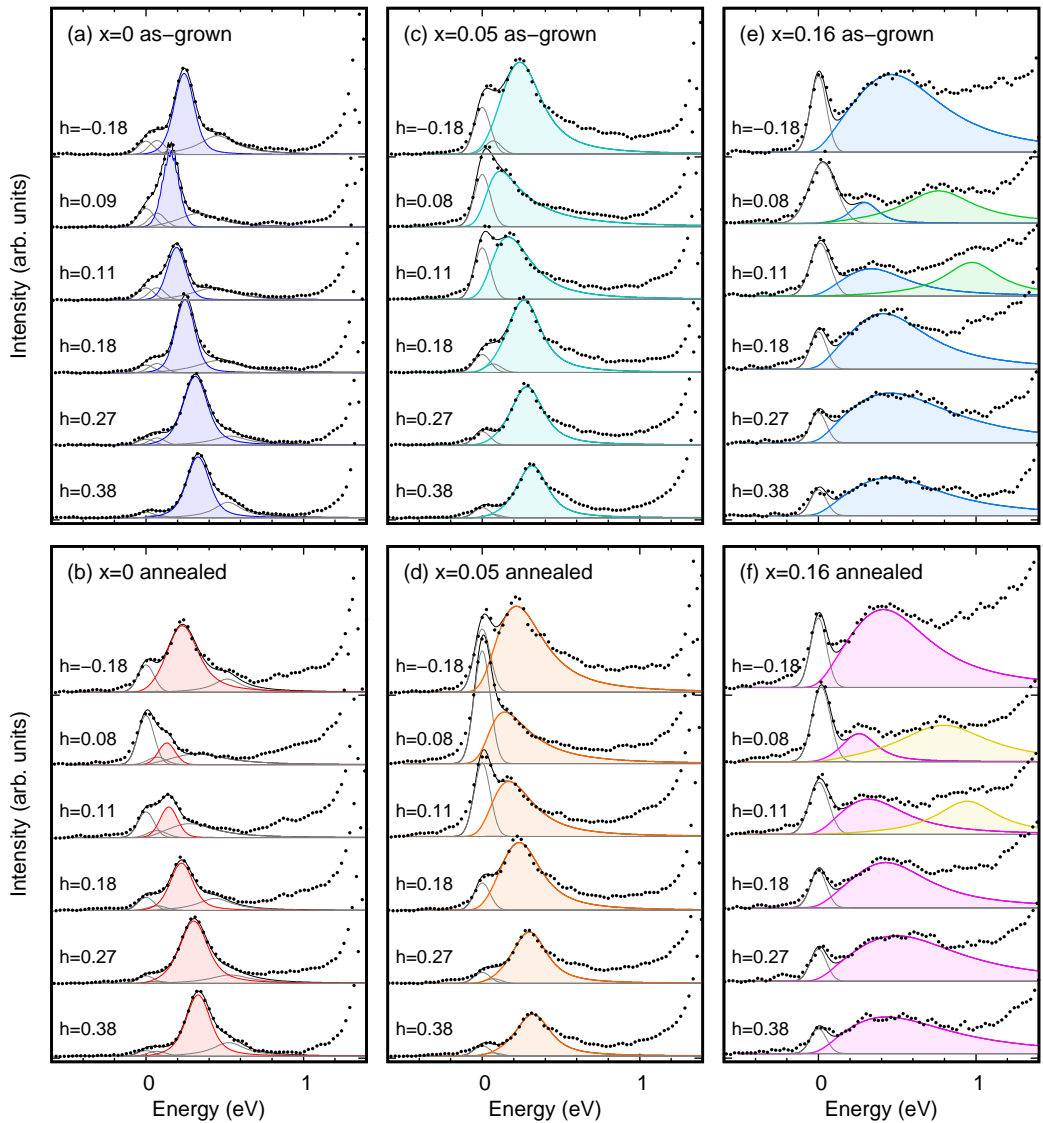}
\caption{(color online). (a)-(f) Fit results of Cu $L_3$-edge RIXS spectra of Nd$_{2-x}$Ce$_x$CuO$_4$.
Dots indicate the data, and lines are components of the fits, as described in the text.
Shaded areas represent the spectral shape of spin and charge excitations.}
\label{fig:cu-l3-fit}
\end{figure*}

\begin{figure*}
\centering
\includegraphics[scale=1.0]{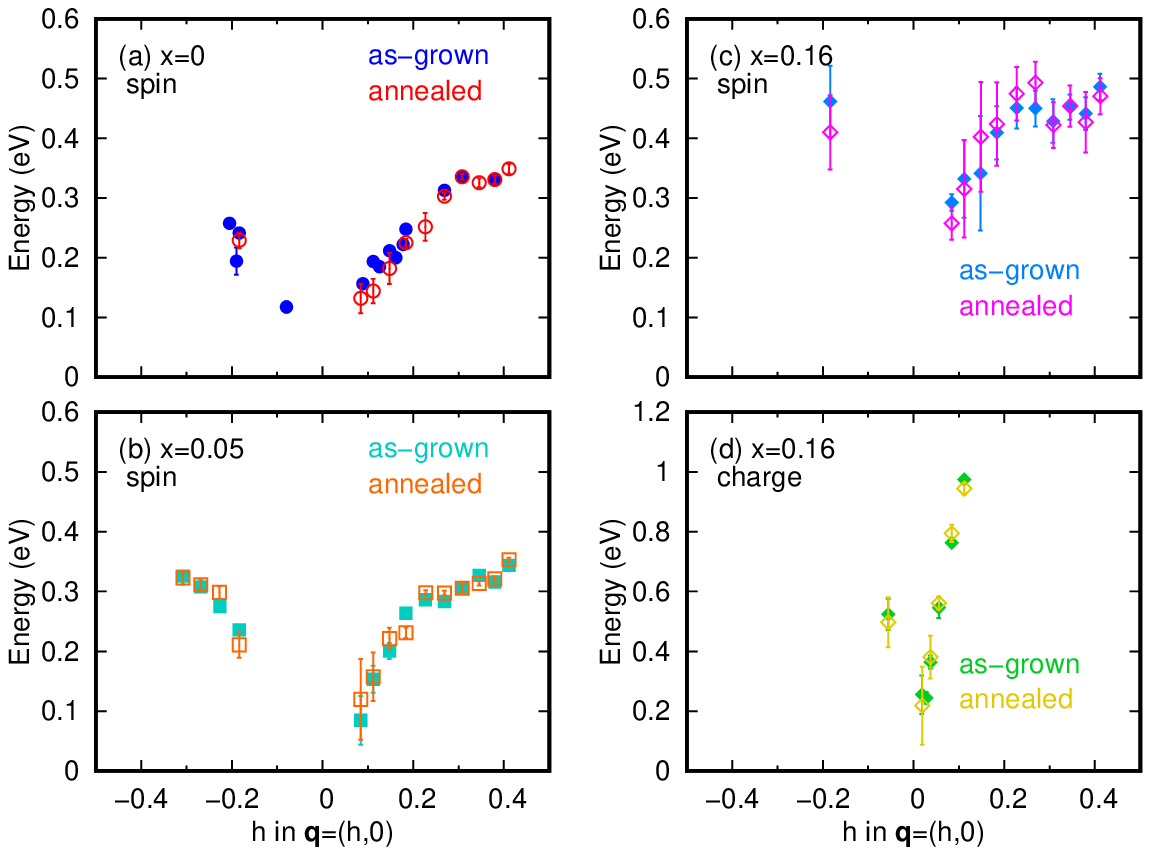}
\caption{(color online). The peak positions of (a-c) spin and (d) charge excitations obtained from fits, as described in the text.
Open and filled symbols indicate as-grown and annealed crystals, respectively.}
\label{fig:cu-l3-qdep}
\end{figure*}

In order to compare the peak positions, we fitted the low-energy part of the spectra to the sum of resolution-limited elastic scattering and a spin-excitation contribution that was modeled as a damped harmonic oscillator multiplied by the Bose-Factor \cite{Monney2016}.
For $x=0$, a low-energy phonon component and two-magnon excitations were included in the fit.
We also analyzed the charge excitations in the $x=0.16$ spectra at low $q$.
Fit results are presented in Fig.~\ref{fig:cu-l3-fit}.
The peak positions obtained from this analysis are summarized in Figs.~\ref{fig:cu-l3-qdep}.
We find that, within the accuracy of the experiment, the peak positions of the spin and charge excitations do not change as a result of the anneal.
The likely reason for the absence of a significant effect on the spin excitations is that, even though an increase in the doped electron concentration is indicated by the Cu $K$-edge RIXS data, the change in carrier density is too small.

\section{Discussion}
Our results demonstrate that the charge excitations in Nd$_{2-x}$Ce$_x$CuO$_4$ are affected by the post-growth anneal.
They indicate that additional electrons are introduced into the CuO$_2$ planes in the annealed crystals, consistent with the basic expectation for oxygen reduction.
While this effect is in line with conventional understanding, the results give some new insights into the electronic structure of the $T'$-structured cuprates.

The Cu $K$-edge RIXS data in Fig.~\ref{fig:cu-k}(a) confirm that intraband excitations associated with doped carriers emerge in the annealed parent compound ($x=0$) and that the carriers are electrons.
In the Cu $K$-edge RIXS data, the strength and the type of the observed excitations vary with the choice of the incident photon energy, and the intraband excitations resonate when a core hole is created at the doped site \cite{Lu2005,Ishii2005b,Jia2012,Wakimoto2013}.
Because the core hole has positive charge, the resonance condition depends on the sign of the charge of the doped carriers.
For the electron-doped cuprates, this resonance energy is slightly below the well-screened intermediate state and close to the absorption edge.
The incident photon energy of 8991 eV chosen in the present study satisfies this condition for incident polarization parallel to the $ab$-plane, and therefore the intraband excitations in the annealed $x=0$ crystal originate from doped electrons.
On the other hand, the resonance condition for the hole-doped cuprates lies above the poorly-screened intermediate state, and it is 9003 eV in La$_{2-x}$Sr$_x$CuO$_4$ \cite{Wakimoto2013}.
It has been established that superconductivity in the nominally electron-doped cuprates is ultimately driven by the formation of hole pockets \cite{Dagan2007,Li2019}.
These hole states are not seen in the present experiment, but should in principle be accessible by tuning to the appropriate resonant condition above the poorly-screened intermediate state.

The observed electron doping of the $x=0$ parent compound indicates that the annealing effect is not limited to the removal of excess apical oxygens.
The oxygen concentration in the reduced sample is likely smaller than 4.0 to preserve charge neutrality, namely, the CuO$_2$ planes and/or Nd-O layers are likely oxygen deficient, as indicated in a structural study \cite{Radaelli1994}.
Electron doping of the $T'$-structured compounds without Ce substitution has been reported for thin films \cite{Yu2007,Horio2018,Horio2018b} and polycrystalline samples \cite{Asano2018,Lin2019,Asano2020}.
Our result constitutes spectroscopic evidence of this effect in a Ce-free single crystal, even though the implied density of doped electrons is relatively small.
While thin films and polycrystalline sample become superconducting without Ce substitution, so far this has not been reported for bulk single crystals; superconductivity in single crystals has been reported for Ce concentrations as low as $x=0.04$ \cite{Brinkmann1995}, to the best of our knowledge.
Even so, we certainly observe direct evidence for electron doping in our annealed $x=0$ single crystal.
Superconductivity in Ce-free bulk crystals can potentially be realized in the future if electrons can be sufficiently doped by improved synthesis and anneal procedures.

For $x=0.16$, the charge excitations observed via Cu $K$-edge RIXS increase in intensity as a result of the anneal and are prominent on the lower-energy side of the broad peak [Fig.~\ref{fig:cu-k}(c)].
A transport study of Pr$_{2-x}$Ce$_x$CuO$_4$ ($x=0.17$) reported that oxygenation of an annealed, superconducting sample not only changes the carrier density, but also introduces disorder as evidenced by an increased residual resistivity \cite{Higgins2006}.
This is in accordance with the observation that post-growth annealing induces a fractional decomposition to rare-earth oxide \cite{Mang2004} and a concomitant repair of Cu vacancies \cite{Kang2007}.
We speculate that the emergence of the highly mobile carriers in the less disordered, annealed superconducting crystals causes not only the Drude component seen in optical conductivity \cite{Arima1993,Onose1999}, but also the intensity change on the lower-energy side of the broad peak in the Cu $K$-edge RIXS spectra.

In contrast to the intensity change in the Cu $K$-edge data, the dispersion of the charge excitations observed via Cu $L_3$-edge RIXS is almost identical between the as-grown and annealed crystals [Fig.~\ref{fig:cu-l3-qdep}(d)].
The origin of this charge mode has been proposed to be intraband particle-hole excitations \cite{Ishii2014}, a quantum phase distinct from superconductivity \cite{Lee2014} and, more recently, plasmon excitations \cite{Hepting2018,Greco2016,Greco2019,Lin2020}.
For simple metals, the plasmon excitation energy is expected to follow the square root of the electron density \cite{Hepting2018,Lin2020}.
However, our data do not follow this energy dependence: even though the electron concentration increases after the anneal, the energy of the charge excitation is unchanged; 
for example, an increase in electron concentration by 0.05 [Fig.~\ref{fig:cu-k}(d)] would be expected to result in an energy shift of about 0.1 eV at $\bm{q}=(0.08,0)$.
The disagreement points to the need to take into account effects of electronic correlations and/or atomic disorder to achieve a complete understanding of the charge excitations.

Finally, we discuss the high-energy spin excitations for $x=0.16$.
Our result demonstrates that the high-energy spin excitations are robust against post-growth annealing, irrespective of the occurrence of the superconductivity.
Whereas this may be due the lack of a strong doping dependence around this Ce concentration \cite{Ishii2014}, it is in stark contrast the low-energy ($< 20$ meV) antiferromagnetic response.
Neutron scattering studies \cite{Fujita2003,Mang2004a,Motoyama2006,Motoyama2007,Yu2010} demonstrate a qualitative change from long-range order in as-grown crystals to short-ranged spin correlations in annealed, superconducting samples.
Therefore, the spin excitations are separated into energy ranges with and without significant post-growth annealing effects, with the low-energy range encompassing the scale of the superconducting gap \cite{Qazilbash2005,Motoyama2006,Yu2010}.

\section{Summary}
Using RIXS at the Cu $K$- and $L_3$-edges, we have studied the effects of standard post-growth annealing on the charge and spin excitations in the electron-doped cuprate Nd$_{2-x}$Ce$_x$CuO$_4$ ($x=0$, 0.05, and 0.16).
The intensity of the charge excitations observed at the Cu $K$-edge is found to increase after the anneal.
Because the incident photon energy is selected at the resonant condition of electron-doped sites, this increase can be ascribed to additional electron doping achieved via the annealing process.
Electron doping is observed even for an annealed single crystal without substitution of Ce ($x=0$).
With a further improved synthesis and/or annealing procedure, it can be expected that electrons will be sufficiently doped and disorder effects further minimized, which should lead to superconductivity in the Ce-free single crystals, as already achieved in thin films and polycrystalline samples.

In contrast, we find that the high-energy spin excitations do not change after the annealing step, except for a slight broadening of the spectrum for $x=0$.
Notably, for $x=0.16$, we find a remarkably close similarity between the spin excitations of as-grown non-superconducting and annealed superconducting crystals, in contrast to previous observations from neutron scattering measurements of significant (qualitative) changes of the low-energy magnetic response.
The effects of post-growth annealing on the spin excitations are therefore limited to relatively low energies comparable to the scale of the superconducting gap.

\begin{acknowledgments}
We thank Amol Singh for technical help during our RIXS measurements at TLS, and Damjan Pelc and Yangmu Li for comments on the manuscript.
This work was performed under the Inter-University Cooperative Research Program of the Institute for Materials Research, Tohoku University (Proposal No.~18K0014).
The work at the University of Minnesota was funded by the Department of Energy through the University of Minnesota Center for Quantum Materials, under Grant No.~DE-SC-0016371.
The synchrotron radiation experiments at SPring-8 were carried out at the BL11XU with the approval of the Japan Synchrotron Radiation Research Institute (JASRI) (Proposals No.~2017B3552 and No.~2018A3555) and those at Taiwan Light Source (TLS) of the National Synchrotron Radiation Research Center (NSRRC) were conducted at beam line 05A1 (Proposals No.~2017-2-120-1 and No.~2017-2-120-9).
This work was financially supported by JSPS KAKENHI Grants No.~16H04004.
\end{acknowledgments}

\bibliography{ncco-an}

\begin{thebibliography}{66}%
\makeatletter
\providecommand \@ifxundefined [1]{%
 \@ifx{#1\undefined}
}%
\providecommand \@ifnum [1]{%
 \ifnum #1\expandafter \@firstoftwo
 \else \expandafter \@secondoftwo
 \fi
}%
\providecommand \@ifx [1]{%
 \ifx #1\expandafter \@firstoftwo
 \else \expandafter \@secondoftwo
 \fi
}%
\providecommand \natexlab [1]{#1}%
\providecommand \enquote  [1]{``#1''}%
\providecommand \bibnamefont  [1]{#1}%
\providecommand \bibfnamefont [1]{#1}%
\providecommand \citenamefont [1]{#1}%
\providecommand \href@noop [0]{\@secondoftwo}%
\providecommand \href [0]{\begingroup \@sanitize@url \@href}%
\providecommand \@href[1]{\@@startlink{#1}\@@href}%
\providecommand \@@href[1]{\endgroup#1\@@endlink}%
\providecommand \@sanitize@url [0]{\catcode `\\12\catcode `\$12\catcode
  `\&12\catcode `\#12\catcode `\^12\catcode `\_12\catcode `\%12\relax}%
\providecommand \@@startlink[1]{}%
\providecommand \@@endlink[0]{}%
\providecommand \url  [0]{\begingroup\@sanitize@url \@url }%
\providecommand \@url [1]{\endgroup\@href {#1}{\urlprefix }}%
\providecommand \urlprefix  [0]{URL }%
\providecommand \Eprint [0]{\href }%
\providecommand \doibase [0]{https://doi.org/}%
\providecommand \selectlanguage [0]{\@gobble}%
\providecommand \bibinfo  [0]{\@secondoftwo}%
\providecommand \bibfield  [0]{\@secondoftwo}%
\providecommand \translation [1]{[#1]}%
\providecommand \BibitemOpen [0]{}%
\providecommand \bibitemStop [0]{}%
\providecommand \bibitemNoStop [0]{.\EOS\space}%
\providecommand \EOS [0]{\spacefactor3000\relax}%
\providecommand \BibitemShut  [1]{\csname bibitem#1\endcsname}%
\let\auto@bib@innerbib\@empty
\bibitem [{\citenamefont {Tokura}\ \emph {et~al.}(1989)\citenamefont {Tokura},
  \citenamefont {Takagi},\ and\ \citenamefont {Uchida}}]{Tokura1989}%
  \BibitemOpen
  \bibfield  {author} {\bibinfo {author} {\bibfnamefont {Y.}~\bibnamefont
  {Tokura}}, \bibinfo {author} {\bibfnamefont {H.}~\bibnamefont {Takagi}},\
  and\ \bibinfo {author} {\bibfnamefont {S.}~\bibnamefont {Uchida}},\ }\href
  {http://dx.doi.org/10.1038/337345a0} {\bibfield  {journal} {\bibinfo
  {journal} {Nature}\ }\textbf {\bibinfo {volume} {337}},\ \bibinfo {pages}
  {345} (\bibinfo {year} {1989})}\BibitemShut {NoStop}%
\bibitem [{\citenamefont {Takagi}\ \emph {et~al.}(1989)\citenamefont {Takagi},
  \citenamefont {Uchida},\ and\ \citenamefont {Tokura}}]{Takagi1989a}%
  \BibitemOpen
  \bibfield  {author} {\bibinfo {author} {\bibfnamefont {H.}~\bibnamefont
  {Takagi}}, \bibinfo {author} {\bibfnamefont {S.}~\bibnamefont {Uchida}},\
  and\ \bibinfo {author} {\bibfnamefont {Y.}~\bibnamefont {Tokura}},\ }\href
  {https://doi.org/10.1103/PhysRevLett.62.1197} {\bibfield  {journal} {\bibinfo
   {journal} {Phys. Rev. Lett.}\ }\textbf {\bibinfo {volume} {62}},\ \bibinfo
  {pages} {1197} (\bibinfo {year} {1989})}\BibitemShut {NoStop}%
\bibitem [{\citenamefont {Matsuda}\ \emph {et~al.}(1992)\citenamefont
  {Matsuda}, \citenamefont {Endoh}, \citenamefont {Yamada}, \citenamefont
  {Kojima}, \citenamefont {Tanaka}, \citenamefont {Birgeneau}, \citenamefont
  {Kastner},\ and\ \citenamefont {Shirane}}]{Matsuda1992}%
  \BibitemOpen
  \bibfield  {author} {\bibinfo {author} {\bibfnamefont {M.}~\bibnamefont
  {Matsuda}}, \bibinfo {author} {\bibfnamefont {Y.}~\bibnamefont {Endoh}},
  \bibinfo {author} {\bibfnamefont {K.}~\bibnamefont {Yamada}}, \bibinfo
  {author} {\bibfnamefont {H.}~\bibnamefont {Kojima}}, \bibinfo {author}
  {\bibfnamefont {I.}~\bibnamefont {Tanaka}}, \bibinfo {author} {\bibfnamefont
  {R.~J.}\ \bibnamefont {Birgeneau}}, \bibinfo {author} {\bibfnamefont {M.~A.}\
  \bibnamefont {Kastner}},\ and\ \bibinfo {author} {\bibfnamefont
  {G.}~\bibnamefont {Shirane}},\ }\href
  {https://doi.org/10.1103/PhysRevB.45.12548} {\bibfield  {journal} {\bibinfo
  {journal} {Phys. Rev. B}\ }\textbf {\bibinfo {volume} {45}},\ \bibinfo
  {pages} {12548} (\bibinfo {year} {1992})}\BibitemShut {NoStop}%
\bibitem [{\citenamefont {Radaelli}\ \emph {et~al.}(1994)\citenamefont
  {Radaelli}, \citenamefont {Jorgensen}, \citenamefont {Schultz}, \citenamefont
  {Peng},\ and\ \citenamefont {Greene}}]{Radaelli1994}%
  \BibitemOpen
  \bibfield  {author} {\bibinfo {author} {\bibfnamefont {P.~G.}\ \bibnamefont
  {Radaelli}}, \bibinfo {author} {\bibfnamefont {J.~D.}\ \bibnamefont
  {Jorgensen}}, \bibinfo {author} {\bibfnamefont {A.~J.}\ \bibnamefont
  {Schultz}}, \bibinfo {author} {\bibfnamefont {J.~L.}\ \bibnamefont {Peng}},\
  and\ \bibinfo {author} {\bibfnamefont {R.~L.}\ \bibnamefont {Greene}},\
  }\href {https://doi.org/10.1103/PhysRevB.49.15322} {\bibfield  {journal}
  {\bibinfo  {journal} {Phys. Rev. B}\ }\textbf {\bibinfo {volume} {49}},\
  \bibinfo {pages} {15322} (\bibinfo {year} {1994})}\BibitemShut {NoStop}%
\bibitem [{\citenamefont {Schultz}\ \emph {et~al.}(1996)\citenamefont
  {Schultz}, \citenamefont {Jorgensen}, \citenamefont {Peng},\ and\
  \citenamefont {Greene}}]{Schultz1996}%
  \BibitemOpen
  \bibfield  {author} {\bibinfo {author} {\bibfnamefont {A.~J.}\ \bibnamefont
  {Schultz}}, \bibinfo {author} {\bibfnamefont {J.~D.}\ \bibnamefont
  {Jorgensen}}, \bibinfo {author} {\bibfnamefont {J.~L.}\ \bibnamefont
  {Peng}},\ and\ \bibinfo {author} {\bibfnamefont {R.~L.}\ \bibnamefont
  {Greene}},\ }\href {https://doi.org/10.1103/PhysRevB.53.5157} {\bibfield
  {journal} {\bibinfo  {journal} {Phys. Rev. B}\ }\textbf {\bibinfo {volume}
  {53}},\ \bibinfo {pages} {5157} (\bibinfo {year} {1996})}\BibitemShut
  {NoStop}%
\bibitem [{\citenamefont {Onose}\ \emph {et~al.}(1999)\citenamefont {Onose},
  \citenamefont {Taguchi}, \citenamefont {Ishikawa}, \citenamefont {Shinomori},
  \citenamefont {Ishizaka},\ and\ \citenamefont {Tokura}}]{Onose1999}%
  \BibitemOpen
  \bibfield  {author} {\bibinfo {author} {\bibfnamefont {Y.}~\bibnamefont
  {Onose}}, \bibinfo {author} {\bibfnamefont {Y.}~\bibnamefont {Taguchi}},
  \bibinfo {author} {\bibfnamefont {T.}~\bibnamefont {Ishikawa}}, \bibinfo
  {author} {\bibfnamefont {S.}~\bibnamefont {Shinomori}}, \bibinfo {author}
  {\bibfnamefont {K.}~\bibnamefont {Ishizaka}},\ and\ \bibinfo {author}
  {\bibfnamefont {Y.}~\bibnamefont {Tokura}},\ }\href
  {https://doi.org/10.1103/PhysRevLett.82.5120} {\bibfield  {journal} {\bibinfo
   {journal} {Phys. Rev. Lett.}\ }\textbf {\bibinfo {volume} {82}},\ \bibinfo
  {pages} {5120} (\bibinfo {year} {1999})}\BibitemShut {NoStop}%
\bibitem [{\citenamefont {Mang}\ \emph
  {et~al.}(2004{\natexlab{a}})\citenamefont {Mang}, \citenamefont {Larochelle},
  \citenamefont {Mehta}, \citenamefont {Vajk}, \citenamefont {Erickson},
  \citenamefont {Lu}, \citenamefont {Buyers}, \citenamefont {Marshall},
  \citenamefont {Prokes},\ and\ \citenamefont {Greven}}]{Mang2004}%
  \BibitemOpen
  \bibfield  {author} {\bibinfo {author} {\bibfnamefont {P.~K.}\ \bibnamefont
  {Mang}}, \bibinfo {author} {\bibfnamefont {S.}~\bibnamefont {Larochelle}},
  \bibinfo {author} {\bibfnamefont {A.}~\bibnamefont {Mehta}}, \bibinfo
  {author} {\bibfnamefont {O.~P.}\ \bibnamefont {Vajk}}, \bibinfo {author}
  {\bibfnamefont {A.~S.}\ \bibnamefont {Erickson}}, \bibinfo {author}
  {\bibfnamefont {L.}~\bibnamefont {Lu}}, \bibinfo {author} {\bibfnamefont
  {W.~J.~L.}\ \bibnamefont {Buyers}}, \bibinfo {author} {\bibfnamefont {A.~F.}\
  \bibnamefont {Marshall}}, \bibinfo {author} {\bibfnamefont {K.}~\bibnamefont
  {Prokes}},\ and\ \bibinfo {author} {\bibfnamefont {M.}~\bibnamefont
  {Greven}},\ }\href {https://doi.org/10.1103/PhysRevB.70.094507} {\bibfield
  {journal} {\bibinfo  {journal} {Phys. Rev. B}\ }\textbf {\bibinfo {volume}
  {70}},\ \bibinfo {pages} {094507} (\bibinfo {year}
  {2004}{\natexlab{a}})}\BibitemShut {NoStop}%
\bibitem [{\citenamefont {Kang}\ \emph {et~al.}(2007)\citenamefont {Kang},
  \citenamefont {Dai}, \citenamefont {Campbell}, \citenamefont {Chupas},
  \citenamefont {Rosenkranz}, \citenamefont {Lee}, \citenamefont {Huang},
  \citenamefont {Li}, \citenamefont {Komiya},\ and\ \citenamefont
  {Ando}}]{Kang2007}%
  \BibitemOpen
  \bibfield  {author} {\bibinfo {author} {\bibfnamefont {H.~J.}\ \bibnamefont
  {Kang}}, \bibinfo {author} {\bibfnamefont {P.}~\bibnamefont {Dai}}, \bibinfo
  {author} {\bibfnamefont {B.~J.}\ \bibnamefont {Campbell}}, \bibinfo {author}
  {\bibfnamefont {P.~J.}\ \bibnamefont {Chupas}}, \bibinfo {author}
  {\bibfnamefont {S.}~\bibnamefont {Rosenkranz}}, \bibinfo {author}
  {\bibfnamefont {P.~L.}\ \bibnamefont {Lee}}, \bibinfo {author} {\bibfnamefont
  {Q.}~\bibnamefont {Huang}}, \bibinfo {author} {\bibfnamefont
  {S.}~\bibnamefont {Li}}, \bibinfo {author} {\bibfnamefont {S.}~\bibnamefont
  {Komiya}},\ and\ \bibinfo {author} {\bibfnamefont {Y.}~\bibnamefont {Ando}},\
  }\href {https://doi.org/10.1038/nmat1847} {\bibfield  {journal} {\bibinfo
  {journal} {Nat. Mater.}\ }\textbf {\bibinfo {volume} {6}},\ \bibinfo {pages}
  {224} (\bibinfo {year} {2007})}\BibitemShut {NoStop}%
\bibitem [{\citenamefont {Armitage}\ \emph {et~al.}(2010)\citenamefont
  {Armitage}, \citenamefont {Fournier},\ and\ \citenamefont
  {Greene}}]{Armitage2010}%
  \BibitemOpen
  \bibfield  {author} {\bibinfo {author} {\bibfnamefont {N.~P.}\ \bibnamefont
  {Armitage}}, \bibinfo {author} {\bibfnamefont {P.}~\bibnamefont {Fournier}},\
  and\ \bibinfo {author} {\bibfnamefont {R.~L.}\ \bibnamefont {Greene}},\
  }\href {https://doi.org/10.1103/RevModPhys.82.2421} {\bibfield  {journal}
  {\bibinfo  {journal} {Rev. Mod. Phys.}\ }\textbf {\bibinfo {volume} {82}},\
  \bibinfo {pages} {2421} (\bibinfo {year} {2010})}\BibitemShut {NoStop}%
\bibitem [{\citenamefont {Motoyama}\ \emph {et~al.}(2007)\citenamefont
  {Motoyama}, \citenamefont {Yu}, \citenamefont {Vishik}, \citenamefont {Vajk},
  \citenamefont {Mang},\ and\ \citenamefont {Greven}}]{Motoyama2007}%
  \BibitemOpen
  \bibfield  {author} {\bibinfo {author} {\bibfnamefont {E.~M.}\ \bibnamefont
  {Motoyama}}, \bibinfo {author} {\bibfnamefont {G.}~\bibnamefont {Yu}},
  \bibinfo {author} {\bibfnamefont {I.~M.}\ \bibnamefont {Vishik}}, \bibinfo
  {author} {\bibfnamefont {O.~P.}\ \bibnamefont {Vajk}}, \bibinfo {author}
  {\bibfnamefont {P.~K.}\ \bibnamefont {Mang}},\ and\ \bibinfo {author}
  {\bibfnamefont {M.}~\bibnamefont {Greven}},\ }\href
  {http://dx.doi.org/10.1038/nature05437} {\bibfield  {journal} {\bibinfo
  {journal} {Nature}\ }\textbf {\bibinfo {volume} {445}},\ \bibinfo {pages}
  {186} (\bibinfo {year} {2007})}\BibitemShut {NoStop}%
\bibitem [{\citenamefont {Brinkmann}\ \emph {et~al.}(1995)\citenamefont
  {Brinkmann}, \citenamefont {Rex}, \citenamefont {Bach},\ and\ \citenamefont
  {Westerholt}}]{Brinkmann1995}%
  \BibitemOpen
  \bibfield  {author} {\bibinfo {author} {\bibfnamefont {M.}~\bibnamefont
  {Brinkmann}}, \bibinfo {author} {\bibfnamefont {T.}~\bibnamefont {Rex}},
  \bibinfo {author} {\bibfnamefont {H.}~\bibnamefont {Bach}},\ and\ \bibinfo
  {author} {\bibfnamefont {K.}~\bibnamefont {Westerholt}},\ }\href
  {https://doi.org/10.1103/PhysRevLett.74.4927} {\bibfield  {journal} {\bibinfo
   {journal} {Phys. Rev. Lett.}\ }\textbf {\bibinfo {volume} {74}},\ \bibinfo
  {pages} {4927} (\bibinfo {year} {1995})}\BibitemShut {NoStop}%
\bibitem [{\citenamefont {Adachi}\ \emph {et~al.}(2013)\citenamefont {Adachi},
  \citenamefont {Mori}, \citenamefont {Takahashi}, \citenamefont {Kato},
  \citenamefont {Nishizaki}, \citenamefont {Sasaki}, \citenamefont
  {Kobayashi},\ and\ \citenamefont {Koike}}]{Adachi2013}%
  \BibitemOpen
  \bibfield  {author} {\bibinfo {author} {\bibfnamefont {T.}~\bibnamefont
  {Adachi}}, \bibinfo {author} {\bibfnamefont {Y.}~\bibnamefont {Mori}},
  \bibinfo {author} {\bibfnamefont {A.}~\bibnamefont {Takahashi}}, \bibinfo
  {author} {\bibfnamefont {M.}~\bibnamefont {Kato}}, \bibinfo {author}
  {\bibfnamefont {T.}~\bibnamefont {Nishizaki}}, \bibinfo {author}
  {\bibfnamefont {T.}~\bibnamefont {Sasaki}}, \bibinfo {author} {\bibfnamefont
  {N.}~\bibnamefont {Kobayashi}},\ and\ \bibinfo {author} {\bibfnamefont
  {Y.}~\bibnamefont {Koike}},\ }\href {https://doi.org/10.7566/JPSJ.82.063713}
  {\bibfield  {journal} {\bibinfo  {journal} {J. Phys. Soc. Jpn.}\ }\textbf
  {\bibinfo {volume} {82}},\ \bibinfo {pages} {063713} (\bibinfo {year}
  {2013})}\BibitemShut {NoStop}%
\bibitem [{\citenamefont {Tsukada}\ \emph {et~al.}(2005)\citenamefont
  {Tsukada}, \citenamefont {Krockenberger}, \citenamefont {Noda}, \citenamefont
  {Yamamoto}, \citenamefont {Manske}, \citenamefont {Alff},\ and\ \citenamefont
  {Naito}}]{Tsukada2005}%
  \BibitemOpen
  \bibfield  {author} {\bibinfo {author} {\bibfnamefont {A.}~\bibnamefont
  {Tsukada}}, \bibinfo {author} {\bibfnamefont {Y.}~\bibnamefont
  {Krockenberger}}, \bibinfo {author} {\bibfnamefont {M.}~\bibnamefont {Noda}},
  \bibinfo {author} {\bibfnamefont {H.}~\bibnamefont {Yamamoto}}, \bibinfo
  {author} {\bibfnamefont {D.}~\bibnamefont {Manske}}, \bibinfo {author}
  {\bibfnamefont {L.}~\bibnamefont {Alff}},\ and\ \bibinfo {author}
  {\bibfnamefont {M.}~\bibnamefont {Naito}},\ }\href
  {https://doi.org/http://dx.doi.org/10.1016/j.ssc.2004.12.011} {\bibfield
  {journal} {\bibinfo  {journal} {Solid State Comm.}\ }\textbf {\bibinfo
  {volume} {133}},\ \bibinfo {pages} {427 } (\bibinfo {year}
  {2005})}\BibitemShut {NoStop}%
\bibitem [{\citenamefont {Takamatsu}\ \emph {et~al.}(2012)\citenamefont
  {Takamatsu}, \citenamefont {Kato}, \citenamefont {Noji},\ and\ \citenamefont
  {Koike}}]{Takamatsu2012}%
  \BibitemOpen
  \bibfield  {author} {\bibinfo {author} {\bibfnamefont {T.}~\bibnamefont
  {Takamatsu}}, \bibinfo {author} {\bibfnamefont {M.}~\bibnamefont {Kato}},
  \bibinfo {author} {\bibfnamefont {T.}~\bibnamefont {Noji}},\ and\ \bibinfo
  {author} {\bibfnamefont {Y.}~\bibnamefont {Koike}},\ }\href
  {http://stacks.iop.org/1882-0786/5/i=7/a=073101} {\bibfield  {journal}
  {\bibinfo  {journal} {Appl. Phys. Express}\ }\textbf {\bibinfo {volume}
  {5}},\ \bibinfo {pages} {073101} (\bibinfo {year} {2012})}\BibitemShut
  {NoStop}%
\bibitem [{\citenamefont {Krockenberger}\ \emph {et~al.}(2013)\citenamefont
  {Krockenberger}, \citenamefont {Irie}, \citenamefont {Matsumoto},
  \citenamefont {Yamagami}, \citenamefont {Mitsuhashi}, \citenamefont
  {Tsukada}, \citenamefont {Naito},\ and\ \citenamefont
  {Yamamoto}}]{Krockenberger2013}%
  \BibitemOpen
  \bibfield  {author} {\bibinfo {author} {\bibfnamefont {Y.}~\bibnamefont
  {Krockenberger}}, \bibinfo {author} {\bibfnamefont {H.}~\bibnamefont {Irie}},
  \bibinfo {author} {\bibfnamefont {O.}~\bibnamefont {Matsumoto}}, \bibinfo
  {author} {\bibfnamefont {K.}~\bibnamefont {Yamagami}}, \bibinfo {author}
  {\bibfnamefont {M.}~\bibnamefont {Mitsuhashi}}, \bibinfo {author}
  {\bibfnamefont {A.}~\bibnamefont {Tsukada}}, \bibinfo {author} {\bibfnamefont
  {M.}~\bibnamefont {Naito}},\ and\ \bibinfo {author} {\bibfnamefont
  {H.}~\bibnamefont {Yamamoto}},\ }\href {http://dx.doi.org/10.1038/srep02235}
  {\bibfield  {journal} {\bibinfo  {journal} {Sci. Rep.}\ }\textbf {\bibinfo
  {volume} {3}},\ \bibinfo {pages} {2235} (\bibinfo {year} {2013})}\BibitemShut
  {NoStop}%
\bibitem [{\citenamefont {Adachi}\ \emph {et~al.}(2017)\citenamefont {Adachi},
  \citenamefont {Kawamata},\ and\ \citenamefont {Koike}}]{Adachi2017}%
  \BibitemOpen
  \bibfield  {author} {\bibinfo {author} {\bibfnamefont {T.}~\bibnamefont
  {Adachi}}, \bibinfo {author} {\bibfnamefont {T.}~\bibnamefont {Kawamata}},\
  and\ \bibinfo {author} {\bibfnamefont {Y.}~\bibnamefont {Koike}},\ }\href
  {https://doi.org/10.3390/condmat2030023} {\bibfield  {journal} {\bibinfo
  {journal} {Condens. Matter}\ }\textbf {\bibinfo {volume} {2}},\ \bibinfo
  {pages} {23} (\bibinfo {year} {2017})}\BibitemShut {NoStop}%
\bibitem [{\citenamefont {Song}\ \emph {et~al.}(2017)\citenamefont {Song},
  \citenamefont {Han}, \citenamefont {Kyung}, \citenamefont {Seo},
  \citenamefont {Cho}, \citenamefont {Kim}, \citenamefont {Arita},
  \citenamefont {Shimada}, \citenamefont {Namatame}, \citenamefont {Taniguchi},
  \citenamefont {Yoshida}, \citenamefont {Eisaki}, \citenamefont {Park},\ and\
  \citenamefont {Kim}}]{Song2017}%
  \BibitemOpen
  \bibfield  {author} {\bibinfo {author} {\bibfnamefont {D.}~\bibnamefont
  {Song}}, \bibinfo {author} {\bibfnamefont {G.}~\bibnamefont {Han}}, \bibinfo
  {author} {\bibfnamefont {W.}~\bibnamefont {Kyung}}, \bibinfo {author}
  {\bibfnamefont {J.}~\bibnamefont {Seo}}, \bibinfo {author} {\bibfnamefont
  {S.}~\bibnamefont {Cho}}, \bibinfo {author} {\bibfnamefont {B.~S.}\
  \bibnamefont {Kim}}, \bibinfo {author} {\bibfnamefont {M.}~\bibnamefont
  {Arita}}, \bibinfo {author} {\bibfnamefont {K.}~\bibnamefont {Shimada}},
  \bibinfo {author} {\bibfnamefont {H.}~\bibnamefont {Namatame}}, \bibinfo
  {author} {\bibfnamefont {M.}~\bibnamefont {Taniguchi}}, \bibinfo {author}
  {\bibfnamefont {Y.}~\bibnamefont {Yoshida}}, \bibinfo {author} {\bibfnamefont
  {H.}~\bibnamefont {Eisaki}}, \bibinfo {author} {\bibfnamefont {S.~R.}\
  \bibnamefont {Park}},\ and\ \bibinfo {author} {\bibfnamefont
  {C.}~\bibnamefont {Kim}},\ }\href
  {https://doi.org/10.1103/PhysRevLett.118.137001} {\bibfield  {journal}
  {\bibinfo  {journal} {Phys. Rev. Lett.}\ }\textbf {\bibinfo {volume} {118}},\
  \bibinfo {pages} {137001} (\bibinfo {year} {2017})}\BibitemShut {NoStop}%
\bibitem [{\citenamefont {Adachi}\ \emph {et~al.}(2016)\citenamefont {Adachi},
  \citenamefont {Takahashi}, \citenamefont {Suzuki}, \citenamefont {Baqiya},
  \citenamefont {Konno}, \citenamefont {Takamatsu}, \citenamefont {Kato},
  \citenamefont {Watanabe}, \citenamefont {Koda}, \citenamefont {Miyazaki},
  \citenamefont {Kadono},\ and\ \citenamefont {Koike}}]{Adachi2016}%
  \BibitemOpen
  \bibfield  {author} {\bibinfo {author} {\bibfnamefont {T.}~\bibnamefont
  {Adachi}}, \bibinfo {author} {\bibfnamefont {A.}~\bibnamefont {Takahashi}},
  \bibinfo {author} {\bibfnamefont {K.~M.}\ \bibnamefont {Suzuki}}, \bibinfo
  {author} {\bibfnamefont {M.~A.}\ \bibnamefont {Baqiya}}, \bibinfo {author}
  {\bibfnamefont {T.}~\bibnamefont {Konno}}, \bibinfo {author} {\bibfnamefont
  {T.}~\bibnamefont {Takamatsu}}, \bibinfo {author} {\bibfnamefont
  {M.}~\bibnamefont {Kato}}, \bibinfo {author} {\bibfnamefont {I.}~\bibnamefont
  {Watanabe}}, \bibinfo {author} {\bibfnamefont {A.}~\bibnamefont {Koda}},
  \bibinfo {author} {\bibfnamefont {M.}~\bibnamefont {Miyazaki}}, \bibinfo
  {author} {\bibfnamefont {R.}~\bibnamefont {Kadono}},\ and\ \bibinfo {author}
  {\bibfnamefont {Y.}~\bibnamefont {Koike}},\ }\href
  {https://doi.org/10.7566/JPSJ.85.114716} {\bibfield  {journal} {\bibinfo
  {journal} {J. Phys. Soc. Jpn.}\ }\textbf {\bibinfo {volume} {85}},\ \bibinfo
  {pages} {114716} (\bibinfo {year} {2016})}\BibitemShut {NoStop}%
\bibitem [{\citenamefont {Horio}\ \emph {et~al.}(2016)\citenamefont {Horio},
  \citenamefont {Adachi}, \citenamefont {Mori}, \citenamefont {Takahashi},
  \citenamefont {Yoshida}, \citenamefont {Suzuki}, \citenamefont {Ambolode~II},
  \citenamefont {Okazaki}, \citenamefont {Ono}, \citenamefont {Kumigashira},
  \citenamefont {Anzai}, \citenamefont {Arita}, \citenamefont {Namatame},
  \citenamefont {Taniguchi}, \citenamefont {Ootsuki}, \citenamefont {Sawada},
  \citenamefont {Takahashi}, \citenamefont {Mizokawa}, \citenamefont {Koike},\
  and\ \citenamefont {Fujimori}}]{Horio2016}%
  \BibitemOpen
  \bibfield  {author} {\bibinfo {author} {\bibfnamefont {M.}~\bibnamefont
  {Horio}}, \bibinfo {author} {\bibfnamefont {T.}~\bibnamefont {Adachi}},
  \bibinfo {author} {\bibfnamefont {Y.}~\bibnamefont {Mori}}, \bibinfo {author}
  {\bibfnamefont {A.}~\bibnamefont {Takahashi}}, \bibinfo {author}
  {\bibfnamefont {T.}~\bibnamefont {Yoshida}}, \bibinfo {author} {\bibfnamefont
  {H.}~\bibnamefont {Suzuki}}, \bibinfo {author} {\bibfnamefont {L.~C.~C.}\
  \bibnamefont {Ambolode~II}}, \bibinfo {author} {\bibfnamefont
  {K.}~\bibnamefont {Okazaki}}, \bibinfo {author} {\bibfnamefont
  {K.}~\bibnamefont {Ono}}, \bibinfo {author} {\bibfnamefont {H.}~\bibnamefont
  {Kumigashira}}, \bibinfo {author} {\bibfnamefont {H.}~\bibnamefont {Anzai}},
  \bibinfo {author} {\bibfnamefont {M.}~\bibnamefont {Arita}}, \bibinfo
  {author} {\bibfnamefont {H.}~\bibnamefont {Namatame}}, \bibinfo {author}
  {\bibfnamefont {M.}~\bibnamefont {Taniguchi}}, \bibinfo {author}
  {\bibfnamefont {D.}~\bibnamefont {Ootsuki}}, \bibinfo {author} {\bibfnamefont
  {K.}~\bibnamefont {Sawada}}, \bibinfo {author} {\bibfnamefont
  {M.}~\bibnamefont {Takahashi}}, \bibinfo {author} {\bibfnamefont
  {T.}~\bibnamefont {Mizokawa}}, \bibinfo {author} {\bibfnamefont
  {Y.}~\bibnamefont {Koike}},\ and\ \bibinfo {author} {\bibfnamefont
  {A.}~\bibnamefont {Fujimori}},\ }\href
  {http://dx.doi.org/10.1038/ncomms10567} {\bibfield  {journal} {\bibinfo
  {journal} {Nat. Commun.}\ }\textbf {\bibinfo {volume} {7}},\ \bibinfo {pages}
  {10567} (\bibinfo {year} {2016})}\BibitemShut {NoStop}%
\bibitem [{\citenamefont {Armitage}\ \emph {et~al.}(2001)\citenamefont
  {Armitage}, \citenamefont {Lu}, \citenamefont {Kim}, \citenamefont
  {Damascelli}, \citenamefont {Shen}, \citenamefont {Ronning}, \citenamefont
  {Feng}, \citenamefont {Bogdanov}, \citenamefont {Shen}, \citenamefont
  {Onose}, \citenamefont {Taguchi}, \citenamefont {Tokura}, \citenamefont
  {Mang}, \citenamefont {Kaneko},\ and\ \citenamefont {Greven}}]{Armitage2001}%
  \BibitemOpen
  \bibfield  {author} {\bibinfo {author} {\bibfnamefont {N.~P.}\ \bibnamefont
  {Armitage}}, \bibinfo {author} {\bibfnamefont {D.~H.}\ \bibnamefont {Lu}},
  \bibinfo {author} {\bibfnamefont {C.}~\bibnamefont {Kim}}, \bibinfo {author}
  {\bibfnamefont {A.}~\bibnamefont {Damascelli}}, \bibinfo {author}
  {\bibfnamefont {K.~M.}\ \bibnamefont {Shen}}, \bibinfo {author}
  {\bibfnamefont {F.}~\bibnamefont {Ronning}}, \bibinfo {author} {\bibfnamefont
  {D.~L.}\ \bibnamefont {Feng}}, \bibinfo {author} {\bibfnamefont
  {P.}~\bibnamefont {Bogdanov}}, \bibinfo {author} {\bibfnamefont {Z.-X.}\
  \bibnamefont {Shen}}, \bibinfo {author} {\bibfnamefont {Y.}~\bibnamefont
  {Onose}}, \bibinfo {author} {\bibfnamefont {Y.}~\bibnamefont {Taguchi}},
  \bibinfo {author} {\bibfnamefont {Y.}~\bibnamefont {Tokura}}, \bibinfo
  {author} {\bibfnamefont {P.~K.}\ \bibnamefont {Mang}}, \bibinfo {author}
  {\bibfnamefont {N.}~\bibnamefont {Kaneko}},\ and\ \bibinfo {author}
  {\bibfnamefont {M.}~\bibnamefont {Greven}},\ }\href
  {https://doi.org/10.1103/PhysRevLett.87.147003} {\bibfield  {journal}
  {\bibinfo  {journal} {Phys. Rev. Lett.}\ }\textbf {\bibinfo {volume} {87}},\
  \bibinfo {pages} {147003} (\bibinfo {year} {2001})}\BibitemShut {NoStop}%
\bibitem [{\citenamefont {Matsui}\ \emph {et~al.}(2005)\citenamefont {Matsui},
  \citenamefont {Terashima}, \citenamefont {Sato}, \citenamefont {Takahashi},
  \citenamefont {Wang}, \citenamefont {Yang}, \citenamefont {Ding},
  \citenamefont {Uefuji},\ and\ \citenamefont {Yamada}}]{Matsui2005}%
  \BibitemOpen
  \bibfield  {author} {\bibinfo {author} {\bibfnamefont {H.}~\bibnamefont
  {Matsui}}, \bibinfo {author} {\bibfnamefont {K.}~\bibnamefont {Terashima}},
  \bibinfo {author} {\bibfnamefont {T.}~\bibnamefont {Sato}}, \bibinfo {author}
  {\bibfnamefont {T.}~\bibnamefont {Takahashi}}, \bibinfo {author}
  {\bibfnamefont {S.-C.}\ \bibnamefont {Wang}}, \bibinfo {author}
  {\bibfnamefont {H.-B.}\ \bibnamefont {Yang}}, \bibinfo {author}
  {\bibfnamefont {H.}~\bibnamefont {Ding}}, \bibinfo {author} {\bibfnamefont
  {T.}~\bibnamefont {Uefuji}},\ and\ \bibinfo {author} {\bibfnamefont
  {K.}~\bibnamefont {Yamada}},\ }\href
  {https://doi.org/10.1103/PhysRevLett.94.047005} {\bibfield  {journal}
  {\bibinfo  {journal} {Phys. Rev. Lett.}\ }\textbf {\bibinfo {volume} {94}},\
  \bibinfo {pages} {047005} (\bibinfo {year} {2005})}\BibitemShut {NoStop}%
\bibitem [{\citenamefont {Scalapino}(2012)}]{Scalapino2012}%
  \BibitemOpen
  \bibfield  {author} {\bibinfo {author} {\bibfnamefont {D.~J.}\ \bibnamefont
  {Scalapino}},\ }\href {https://doi.org/10.1103/RevModPhys.84.1383} {\bibfield
   {journal} {\bibinfo  {journal} {Rev. Mod. Phys.}\ }\textbf {\bibinfo
  {volume} {84}},\ \bibinfo {pages} {1383} (\bibinfo {year}
  {2012})}\BibitemShut {NoStop}%
\bibitem [{\citenamefont {Castellani}\ \emph {et~al.}(1996)\citenamefont
  {Castellani}, \citenamefont {Di~Castro},\ and\ \citenamefont
  {Grilli}}]{Castellani1996}%
  \BibitemOpen
  \bibfield  {author} {\bibinfo {author} {\bibfnamefont {C.}~\bibnamefont
  {Castellani}}, \bibinfo {author} {\bibfnamefont {C.}~\bibnamefont
  {Di~Castro}},\ and\ \bibinfo {author} {\bibfnamefont {M.}~\bibnamefont
  {Grilli}},\ }\href {https://doi.org/10.1007/s002570050347} {\bibfield
  {journal} {\bibinfo  {journal} {Z. Phys. B}\ }\textbf {\bibinfo {volume}
  {103}},\ \bibinfo {pages} {137} (\bibinfo {year} {1996})}\BibitemShut
  {NoStop}%
\bibitem [{\citenamefont {Kivelson}\ \emph {et~al.}(2003)\citenamefont
  {Kivelson}, \citenamefont {Bindloss}, \citenamefont {Fradkin}, \citenamefont
  {Oganesyan}, \citenamefont {Tranquada}, \citenamefont {Kapitulnik},\ and\
  \citenamefont {Howald}}]{Kivelson2003}%
  \BibitemOpen
  \bibfield  {author} {\bibinfo {author} {\bibfnamefont {S.~A.}\ \bibnamefont
  {Kivelson}}, \bibinfo {author} {\bibfnamefont {I.~P.}\ \bibnamefont
  {Bindloss}}, \bibinfo {author} {\bibfnamefont {E.}~\bibnamefont {Fradkin}},
  \bibinfo {author} {\bibfnamefont {V.}~\bibnamefont {Oganesyan}}, \bibinfo
  {author} {\bibfnamefont {J.~M.}\ \bibnamefont {Tranquada}}, \bibinfo {author}
  {\bibfnamefont {A.}~\bibnamefont {Kapitulnik}},\ and\ \bibinfo {author}
  {\bibfnamefont {C.}~\bibnamefont {Howald}},\ }\href
  {https://doi.org/10.1103/RevModPhys.75.1201} {\bibfield  {journal} {\bibinfo
  {journal} {Rev. Mod. Phys.}\ }\textbf {\bibinfo {volume} {75}},\ \bibinfo
  {pages} {1201} (\bibinfo {year} {2003})}\BibitemShut {NoStop}%
\bibitem [{\citenamefont {Ament}\ \emph {et~al.}(2011)\citenamefont {Ament},
  \citenamefont {van Veenendaal}, \citenamefont {Devereaux}, \citenamefont
  {Hill},\ and\ \citenamefont {van~den Brink}}]{Ament2011}%
  \BibitemOpen
  \bibfield  {author} {\bibinfo {author} {\bibfnamefont {L.~J.~P.}\
  \bibnamefont {Ament}}, \bibinfo {author} {\bibfnamefont {M.}~\bibnamefont
  {van Veenendaal}}, \bibinfo {author} {\bibfnamefont {T.~P.}\ \bibnamefont
  {Devereaux}}, \bibinfo {author} {\bibfnamefont {J.~P.}\ \bibnamefont
  {Hill}},\ and\ \bibinfo {author} {\bibfnamefont {J.}~\bibnamefont {van~den
  Brink}},\ }\href {https://doi.org/10.1103/RevModPhys.83.705} {\bibfield
  {journal} {\bibinfo  {journal} {Rev. Mod. Phys.}\ }\textbf {\bibinfo {volume}
  {83}},\ \bibinfo {pages} {705} (\bibinfo {year} {2011})}\BibitemShut
  {NoStop}%
\bibitem [{\citenamefont {Ishii}\ \emph {et~al.}(2013)\citenamefont {Ishii},
  \citenamefont {Tohyama},\ and\ \citenamefont {Mizuki}}]{Ishii2013}%
  \BibitemOpen
  \bibfield  {author} {\bibinfo {author} {\bibfnamefont {K.}~\bibnamefont
  {Ishii}}, \bibinfo {author} {\bibfnamefont {T.}~\bibnamefont {Tohyama}},\
  and\ \bibinfo {author} {\bibfnamefont {J.}~\bibnamefont {Mizuki}},\ }\href
  {https://doi.org/10.1143/JPSJ.82.021015} {\bibfield  {journal} {\bibinfo
  {journal} {J. Phys. Soc. Jpn.}\ }\textbf {\bibinfo {volume} {82}},\ \bibinfo
  {pages} {021015} (\bibinfo {year} {2013})}\BibitemShut {NoStop}%
\bibitem [{\citenamefont {Dean}(2015)}]{Dean2015}%
  \BibitemOpen
  \bibfield  {author} {\bibinfo {author} {\bibfnamefont {M.~P.~M.}\
  \bibnamefont {Dean}},\ }\href
  {http://www.sciencedirect.com/science/article/pii/S0304885314002868}
  {\bibfield  {journal} {\bibinfo  {journal} {J. Magn. Mag. Mater.}\ }\textbf
  {\bibinfo {volume} {376}},\ \bibinfo {pages} {3} (\bibinfo {year}
  {2015})}\BibitemShut {NoStop}%
\bibitem [{\citenamefont {Ishii}\ \emph {et~al.}(2005)\citenamefont {Ishii},
  \citenamefont {Tsutsui}, \citenamefont {Endoh}, \citenamefont {Tohyama},
  \citenamefont {Maekawa}, \citenamefont {Hoesch}, \citenamefont {Kuzushita},
  \citenamefont {Tsubota}, \citenamefont {Inami}, \citenamefont {Mizuki},
  \citenamefont {Murakami},\ and\ \citenamefont {Yamada}}]{Ishii2005b}%
  \BibitemOpen
  \bibfield  {author} {\bibinfo {author} {\bibfnamefont {K.}~\bibnamefont
  {Ishii}}, \bibinfo {author} {\bibfnamefont {K.}~\bibnamefont {Tsutsui}},
  \bibinfo {author} {\bibfnamefont {Y.}~\bibnamefont {Endoh}}, \bibinfo
  {author} {\bibfnamefont {T.}~\bibnamefont {Tohyama}}, \bibinfo {author}
  {\bibfnamefont {S.}~\bibnamefont {Maekawa}}, \bibinfo {author} {\bibfnamefont
  {M.}~\bibnamefont {Hoesch}}, \bibinfo {author} {\bibfnamefont
  {K.}~\bibnamefont {Kuzushita}}, \bibinfo {author} {\bibfnamefont
  {M.}~\bibnamefont {Tsubota}}, \bibinfo {author} {\bibfnamefont
  {T.}~\bibnamefont {Inami}}, \bibinfo {author} {\bibfnamefont
  {J.}~\bibnamefont {Mizuki}}, \bibinfo {author} {\bibfnamefont
  {Y.}~\bibnamefont {Murakami}},\ and\ \bibinfo {author} {\bibfnamefont
  {K.}~\bibnamefont {Yamada}},\ }\href
  {https://doi.org/10.1103/PhysRevLett.94.207003} {\bibfield  {journal}
  {\bibinfo  {journal} {Phys. Rev. Lett.}\ }\textbf {\bibinfo {volume} {94}},\
  \bibinfo {eid} {207003} (\bibinfo {year} {2005})}\BibitemShut {NoStop}%
\bibitem [{\citenamefont {Ishii}\ \emph {et~al.}(2006)\citenamefont {Ishii},
  \citenamefont {Tsutsui}, \citenamefont {Endoh}, \citenamefont {Tohyama},
  \citenamefont {Maekawa}, \citenamefont {Hoesch}, \citenamefont {Kuzushita},
  \citenamefont {Inami}, \citenamefont {Tsubota}, \citenamefont {Yamada},
  \citenamefont {Murakami},\ and\ \citenamefont {Mizuki}}]{Ishii2006}%
  \BibitemOpen
  \bibfield  {author} {\bibinfo {author} {\bibfnamefont {K.}~\bibnamefont
  {Ishii}}, \bibinfo {author} {\bibfnamefont {K.}~\bibnamefont {Tsutsui}},
  \bibinfo {author} {\bibfnamefont {Y.}~\bibnamefont {Endoh}}, \bibinfo
  {author} {\bibfnamefont {T.}~\bibnamefont {Tohyama}}, \bibinfo {author}
  {\bibfnamefont {S.}~\bibnamefont {Maekawa}}, \bibinfo {author} {\bibfnamefont
  {M.}~\bibnamefont {Hoesch}}, \bibinfo {author} {\bibfnamefont
  {K.}~\bibnamefont {Kuzushita}}, \bibinfo {author} {\bibfnamefont
  {T.}~\bibnamefont {Inami}}, \bibinfo {author} {\bibfnamefont
  {M.}~\bibnamefont {Tsubota}}, \bibinfo {author} {\bibfnamefont
  {K.}~\bibnamefont {Yamada}}, \bibinfo {author} {\bibfnamefont
  {Y.}~\bibnamefont {Murakami}},\ and\ \bibinfo {author} {\bibfnamefont
  {J.}~\bibnamefont {Mizuki}},\ }\href {https://doi.org/10.1063/1.2354756}
  {\bibfield  {journal} {\bibinfo  {journal} {AIP Conf. Proc.}\ }\textbf
  {\bibinfo {volume} {850}},\ \bibinfo {pages} {403} (\bibinfo {year}
  {2006})}\BibitemShut {NoStop}%
\bibitem [{\citenamefont {Li}\ \emph {et~al.}(2008)\citenamefont {Li},
  \citenamefont {Qian}, \citenamefont {Wray}, \citenamefont {Hsieh},
  \citenamefont {Xia}, \citenamefont {Kaga}, \citenamefont {Sasagawa},
  \citenamefont {Takagi}, \citenamefont {Markiewicz}, \citenamefont {Bansil},
  \citenamefont {Eisaki}, \citenamefont {Uchida},\ and\ \citenamefont
  {Hasan}}]{Li2008}%
  \BibitemOpen
  \bibfield  {author} {\bibinfo {author} {\bibfnamefont {Y.~W.}\ \bibnamefont
  {Li}}, \bibinfo {author} {\bibfnamefont {D.}~\bibnamefont {Qian}}, \bibinfo
  {author} {\bibfnamefont {L.}~\bibnamefont {Wray}}, \bibinfo {author}
  {\bibfnamefont {D.}~\bibnamefont {Hsieh}}, \bibinfo {author} {\bibfnamefont
  {Y.}~\bibnamefont {Xia}}, \bibinfo {author} {\bibfnamefont {Y.}~\bibnamefont
  {Kaga}}, \bibinfo {author} {\bibfnamefont {T.}~\bibnamefont {Sasagawa}},
  \bibinfo {author} {\bibfnamefont {H.}~\bibnamefont {Takagi}}, \bibinfo
  {author} {\bibfnamefont {R.~S.}\ \bibnamefont {Markiewicz}}, \bibinfo
  {author} {\bibfnamefont {A.}~\bibnamefont {Bansil}}, \bibinfo {author}
  {\bibfnamefont {H.}~\bibnamefont {Eisaki}}, \bibinfo {author} {\bibfnamefont
  {S.}~\bibnamefont {Uchida}},\ and\ \bibinfo {author} {\bibfnamefont {M.~Z.}\
  \bibnamefont {Hasan}},\ }\href {https://doi.org/10.1103/PhysRevB.78.073104}
  {\bibfield  {journal} {\bibinfo  {journal} {Phys. Rev. B}\ }\textbf {\bibinfo
  {volume} {78}},\ \bibinfo {eid} {073104} (\bibinfo {year}
  {2008})}\BibitemShut {NoStop}%
\bibitem [{\citenamefont {Ishii}\ \emph {et~al.}(2014)\citenamefont {Ishii},
  \citenamefont {Fujita}, \citenamefont {Sasaki}, \citenamefont {Minola},
  \citenamefont {Dellea}, \citenamefont {Mazzoli}, \citenamefont {Kummer},
  \citenamefont {Ghiringhelli}, \citenamefont {Braicovich}, \citenamefont
  {Tohyama}, \citenamefont {Tsutsumi}, \citenamefont {Sato}, \citenamefont
  {Kajimoto}, \citenamefont {Ikeuchi}, \citenamefont {Yamada}, \citenamefont
  {Yoshida}, \citenamefont {Kurooka},\ and\ \citenamefont
  {Mizuki}}]{Ishii2014}%
  \BibitemOpen
  \bibfield  {author} {\bibinfo {author} {\bibfnamefont {K.}~\bibnamefont
  {Ishii}}, \bibinfo {author} {\bibfnamefont {M.}~\bibnamefont {Fujita}},
  \bibinfo {author} {\bibfnamefont {T.}~\bibnamefont {Sasaki}}, \bibinfo
  {author} {\bibfnamefont {M.}~\bibnamefont {Minola}}, \bibinfo {author}
  {\bibfnamefont {G.}~\bibnamefont {Dellea}}, \bibinfo {author} {\bibfnamefont
  {C.}~\bibnamefont {Mazzoli}}, \bibinfo {author} {\bibfnamefont
  {K.}~\bibnamefont {Kummer}}, \bibinfo {author} {\bibfnamefont
  {G.}~\bibnamefont {Ghiringhelli}}, \bibinfo {author} {\bibfnamefont
  {L.}~\bibnamefont {Braicovich}}, \bibinfo {author} {\bibfnamefont
  {T.}~\bibnamefont {Tohyama}}, \bibinfo {author} {\bibfnamefont
  {K.}~\bibnamefont {Tsutsumi}}, \bibinfo {author} {\bibfnamefont
  {K.}~\bibnamefont {Sato}}, \bibinfo {author} {\bibfnamefont {R.}~\bibnamefont
  {Kajimoto}}, \bibinfo {author} {\bibfnamefont {K.}~\bibnamefont {Ikeuchi}},
  \bibinfo {author} {\bibfnamefont {K.}~\bibnamefont {Yamada}}, \bibinfo
  {author} {\bibfnamefont {M.}~\bibnamefont {Yoshida}}, \bibinfo {author}
  {\bibfnamefont {M.}~\bibnamefont {Kurooka}},\ and\ \bibinfo {author}
  {\bibfnamefont {J.}~\bibnamefont {Mizuki}},\ }\href
  {http://dx.doi.org/10.1038/ncomms4714} {\bibfield  {journal} {\bibinfo
  {journal} {Nat. Commun.}\ }\textbf {\bibinfo {volume} {5}},\ \bibinfo {pages}
  {3714} (\bibinfo {year} {2014})}\BibitemShut {NoStop}%
\bibitem [{\citenamefont {Ishii}\ \emph {et~al.}(2019)\citenamefont {Ishii},
  \citenamefont {Kurooka}, \citenamefont {Shimizu}, \citenamefont {Fujita},
  \citenamefont {Yamada},\ and\ \citenamefont {Mizuki}}]{Ishii2019}%
  \BibitemOpen
  \bibfield  {author} {\bibinfo {author} {\bibfnamefont {K.}~\bibnamefont
  {Ishii}}, \bibinfo {author} {\bibfnamefont {M.}~\bibnamefont {Kurooka}},
  \bibinfo {author} {\bibfnamefont {Y.}~\bibnamefont {Shimizu}}, \bibinfo
  {author} {\bibfnamefont {M.}~\bibnamefont {Fujita}}, \bibinfo {author}
  {\bibfnamefont {K.}~\bibnamefont {Yamada}},\ and\ \bibinfo {author}
  {\bibfnamefont {J.}~\bibnamefont {Mizuki}},\ }\href
  {https://doi.org/10.7566/JPSJ.88.075001} {\bibfield  {journal} {\bibinfo
  {journal} {J. Phys. Soc. Jpn.}\ }\textbf {\bibinfo {volume} {88}},\ \bibinfo
  {pages} {075001} (\bibinfo {year} {2019})}\BibitemShut {NoStop}%
\bibitem [{\citenamefont {Lee}\ \emph {et~al.}(2014)\citenamefont {Lee},
  \citenamefont {Lee}, \citenamefont {Nowadnick}, \citenamefont {Gerber},
  \citenamefont {Tabis}, \citenamefont {Huang}, \citenamefont {Strocov},
  \citenamefont {Motoyama}, \citenamefont {Yu}, \citenamefont {Moritz},
  \citenamefont {Huang}, \citenamefont {Wang}, \citenamefont {Huang},
  \citenamefont {Wu}, \citenamefont {Chen}, \citenamefont {Huang},
  \citenamefont {Greven}, \citenamefont {Schmitt}, \citenamefont {Shen},\ and\
  \citenamefont {Devereaux}}]{Lee2014}%
  \BibitemOpen
  \bibfield  {author} {\bibinfo {author} {\bibfnamefont {W.~S.}\ \bibnamefont
  {Lee}}, \bibinfo {author} {\bibfnamefont {J.~J.}\ \bibnamefont {Lee}},
  \bibinfo {author} {\bibfnamefont {E.~A.}\ \bibnamefont {Nowadnick}}, \bibinfo
  {author} {\bibfnamefont {S.}~\bibnamefont {Gerber}}, \bibinfo {author}
  {\bibfnamefont {W.}~\bibnamefont {Tabis}}, \bibinfo {author} {\bibfnamefont
  {S.~W.}\ \bibnamefont {Huang}}, \bibinfo {author} {\bibfnamefont {V.~N.}\
  \bibnamefont {Strocov}}, \bibinfo {author} {\bibfnamefont {E.~M.}\
  \bibnamefont {Motoyama}}, \bibinfo {author} {\bibfnamefont {G.}~\bibnamefont
  {Yu}}, \bibinfo {author} {\bibfnamefont {B.}~\bibnamefont {Moritz}}, \bibinfo
  {author} {\bibfnamefont {H.~Y.}\ \bibnamefont {Huang}}, \bibinfo {author}
  {\bibfnamefont {R.~P.}\ \bibnamefont {Wang}}, \bibinfo {author}
  {\bibfnamefont {Y.~B.}\ \bibnamefont {Huang}}, \bibinfo {author}
  {\bibfnamefont {W.~B.}\ \bibnamefont {Wu}}, \bibinfo {author} {\bibfnamefont
  {C.~T.}\ \bibnamefont {Chen}}, \bibinfo {author} {\bibfnamefont {D.~J.}\
  \bibnamefont {Huang}}, \bibinfo {author} {\bibfnamefont {M.}~\bibnamefont
  {Greven}}, \bibinfo {author} {\bibfnamefont {T.}~\bibnamefont {Schmitt}},
  \bibinfo {author} {\bibfnamefont {Z.~X.}\ \bibnamefont {Shen}},\ and\
  \bibinfo {author} {\bibfnamefont {T.~P.}\ \bibnamefont {Devereaux}},\ }\href
  {http://dx.doi.org/10.1038/nphys3117} {\bibfield  {journal} {\bibinfo
  {journal} {Nat. Phys.}\ }\textbf {\bibinfo {volume} {10}},\ \bibinfo {pages}
  {883} (\bibinfo {year} {2014})}\BibitemShut {NoStop}%
\bibitem [{\citenamefont {da~Silva~Neto}\ \emph {et~al.}(2016)\citenamefont
  {da~Silva~Neto}, \citenamefont {Yu}, \citenamefont {Minola}, \citenamefont
  {Sutarto}, \citenamefont {Schierle}, \citenamefont {Boschini}, \citenamefont
  {Zonno}, \citenamefont {Bluschke}, \citenamefont {Higgins}, \citenamefont
  {Li}, \citenamefont {Yu}, \citenamefont {Weschke}, \citenamefont {He},
  \citenamefont {Le~Tacon}, \citenamefont {Greene}, \citenamefont {Greven},
  \citenamefont {Sawatzky}, \citenamefont {Keimer},\ and\ \citenamefont
  {Damascelli}}]{SilvaNeto2016}%
  \BibitemOpen
  \bibfield  {author} {\bibinfo {author} {\bibfnamefont {E.~H.}\ \bibnamefont
  {da~Silva~Neto}}, \bibinfo {author} {\bibfnamefont {B.}~\bibnamefont {Yu}},
  \bibinfo {author} {\bibfnamefont {M.}~\bibnamefont {Minola}}, \bibinfo
  {author} {\bibfnamefont {R.}~\bibnamefont {Sutarto}}, \bibinfo {author}
  {\bibfnamefont {E.}~\bibnamefont {Schierle}}, \bibinfo {author}
  {\bibfnamefont {F.}~\bibnamefont {Boschini}}, \bibinfo {author}
  {\bibfnamefont {M.}~\bibnamefont {Zonno}}, \bibinfo {author} {\bibfnamefont
  {M.}~\bibnamefont {Bluschke}}, \bibinfo {author} {\bibfnamefont
  {J.}~\bibnamefont {Higgins}}, \bibinfo {author} {\bibfnamefont
  {Y.}~\bibnamefont {Li}}, \bibinfo {author} {\bibfnamefont {G.}~\bibnamefont
  {Yu}}, \bibinfo {author} {\bibfnamefont {E.}~\bibnamefont {Weschke}},
  \bibinfo {author} {\bibfnamefont {F.}~\bibnamefont {He}}, \bibinfo {author}
  {\bibfnamefont {M.}~\bibnamefont {Le~Tacon}}, \bibinfo {author}
  {\bibfnamefont {R.~L.}\ \bibnamefont {Greene}}, \bibinfo {author}
  {\bibfnamefont {M.}~\bibnamefont {Greven}}, \bibinfo {author} {\bibfnamefont
  {G.~A.}\ \bibnamefont {Sawatzky}}, \bibinfo {author} {\bibfnamefont
  {B.}~\bibnamefont {Keimer}},\ and\ \bibinfo {author} {\bibfnamefont
  {A.}~\bibnamefont {Damascelli}},\ }\href
  {https://doi.org/10.1126/sciadv.1600782} {\bibfield  {journal} {\bibinfo
  {journal} {Sci. Adv.}\ }\textbf {\bibinfo {volume} {2}},\ \bibinfo {pages}
  {e1600782} (\bibinfo {year} {2016})}\BibitemShut {NoStop}%
\bibitem [{\citenamefont {Dellea}\ \emph {et~al.}(2017)\citenamefont {Dellea},
  \citenamefont {Minola}, \citenamefont {Galdi}, \citenamefont {Di~Castro},
  \citenamefont {Aruta}, \citenamefont {Brookes}, \citenamefont {Jia},
  \citenamefont {Mazzoli}, \citenamefont {Moretti~Sala}, \citenamefont
  {Moritz}, \citenamefont {Orgiani}, \citenamefont {Schlom}, \citenamefont
  {Tebano}, \citenamefont {Balestrino}, \citenamefont {Braicovich},
  \citenamefont {Devereaux}, \citenamefont {Maritato},\ and\ \citenamefont
  {Ghiringhelli}}]{Dellea2017}%
  \BibitemOpen
  \bibfield  {author} {\bibinfo {author} {\bibfnamefont {G.}~\bibnamefont
  {Dellea}}, \bibinfo {author} {\bibfnamefont {M.}~\bibnamefont {Minola}},
  \bibinfo {author} {\bibfnamefont {A.}~\bibnamefont {Galdi}}, \bibinfo
  {author} {\bibfnamefont {D.}~\bibnamefont {Di~Castro}}, \bibinfo {author}
  {\bibfnamefont {C.}~\bibnamefont {Aruta}}, \bibinfo {author} {\bibfnamefont
  {N.~B.}\ \bibnamefont {Brookes}}, \bibinfo {author} {\bibfnamefont {C.~J.}\
  \bibnamefont {Jia}}, \bibinfo {author} {\bibfnamefont {C.}~\bibnamefont
  {Mazzoli}}, \bibinfo {author} {\bibfnamefont {M.}~\bibnamefont
  {Moretti~Sala}}, \bibinfo {author} {\bibfnamefont {B.}~\bibnamefont
  {Moritz}}, \bibinfo {author} {\bibfnamefont {P.}~\bibnamefont {Orgiani}},
  \bibinfo {author} {\bibfnamefont {D.~G.}\ \bibnamefont {Schlom}}, \bibinfo
  {author} {\bibfnamefont {A.}~\bibnamefont {Tebano}}, \bibinfo {author}
  {\bibfnamefont {G.}~\bibnamefont {Balestrino}}, \bibinfo {author}
  {\bibfnamefont {L.}~\bibnamefont {Braicovich}}, \bibinfo {author}
  {\bibfnamefont {T.~P.}\ \bibnamefont {Devereaux}}, \bibinfo {author}
  {\bibfnamefont {L.}~\bibnamefont {Maritato}},\ and\ \bibinfo {author}
  {\bibfnamefont {G.}~\bibnamefont {Ghiringhelli}},\ }\href
  {https://doi.org/10.1103/PhysRevB.96.115117} {\bibfield  {journal} {\bibinfo
  {journal} {Phys. Rev. B}\ }\textbf {\bibinfo {volume} {96}},\ \bibinfo
  {pages} {115117} (\bibinfo {year} {2017})}\BibitemShut {NoStop}%
\bibitem [{\citenamefont {da~Silva~Neto}\ \emph {et~al.}(2018)\citenamefont
  {da~Silva~Neto}, \citenamefont {Minola}, \citenamefont {Yu}, \citenamefont
  {Tabis}, \citenamefont {Bluschke}, \citenamefont {Unruh}, \citenamefont
  {Suzuki}, \citenamefont {Li}, \citenamefont {Yu}, \citenamefont {Betto},
  \citenamefont {Kummer}, \citenamefont {Yakhou}, \citenamefont {Brookes},
  \citenamefont {Le~Tacon}, \citenamefont {Greven}, \citenamefont {Keimer},\
  and\ \citenamefont {Damascelli}}]{SilvaNeto2018}%
  \BibitemOpen
  \bibfield  {author} {\bibinfo {author} {\bibfnamefont {E.~H.}\ \bibnamefont
  {da~Silva~Neto}}, \bibinfo {author} {\bibfnamefont {M.}~\bibnamefont
  {Minola}}, \bibinfo {author} {\bibfnamefont {B.}~\bibnamefont {Yu}}, \bibinfo
  {author} {\bibfnamefont {W.}~\bibnamefont {Tabis}}, \bibinfo {author}
  {\bibfnamefont {M.}~\bibnamefont {Bluschke}}, \bibinfo {author}
  {\bibfnamefont {D.}~\bibnamefont {Unruh}}, \bibinfo {author} {\bibfnamefont
  {H.}~\bibnamefont {Suzuki}}, \bibinfo {author} {\bibfnamefont
  {Y.}~\bibnamefont {Li}}, \bibinfo {author} {\bibfnamefont {G.}~\bibnamefont
  {Yu}}, \bibinfo {author} {\bibfnamefont {D.}~\bibnamefont {Betto}}, \bibinfo
  {author} {\bibfnamefont {K.}~\bibnamefont {Kummer}}, \bibinfo {author}
  {\bibfnamefont {F.}~\bibnamefont {Yakhou}}, \bibinfo {author} {\bibfnamefont
  {N.~B.}\ \bibnamefont {Brookes}}, \bibinfo {author} {\bibfnamefont
  {M.}~\bibnamefont {Le~Tacon}}, \bibinfo {author} {\bibfnamefont
  {M.}~\bibnamefont {Greven}}, \bibinfo {author} {\bibfnamefont
  {B.}~\bibnamefont {Keimer}},\ and\ \bibinfo {author} {\bibfnamefont
  {A.}~\bibnamefont {Damascelli}},\ }\href
  {https://doi.org/10.1103/PhysRevB.98.161114} {\bibfield  {journal} {\bibinfo
  {journal} {Phys. Rev. B}\ }\textbf {\bibinfo {volume} {98}},\ \bibinfo
  {pages} {161114} (\bibinfo {year} {2018})}\BibitemShut {NoStop}%
\bibitem [{\citenamefont {Hepting}\ \emph {et~al.}(2018)\citenamefont
  {Hepting}, \citenamefont {Chaix}, \citenamefont {Huang}, \citenamefont
  {Fumagalli}, \citenamefont {Peng}, \citenamefont {Moritz}, \citenamefont
  {Kummer}, \citenamefont {Brookes}, \citenamefont {Lee}, \citenamefont
  {Hashimoto}, \citenamefont {Sarkar}, \citenamefont {He}, \citenamefont
  {Rotundu}, \citenamefont {Lee}, \citenamefont {Greene}, \citenamefont
  {Braicovich}, \citenamefont {Ghiringhelli}, \citenamefont {Shen},
  \citenamefont {Devereaux},\ and\ \citenamefont {Lee}}]{Hepting2018}%
  \BibitemOpen
  \bibfield  {author} {\bibinfo {author} {\bibfnamefont {M.}~\bibnamefont
  {Hepting}}, \bibinfo {author} {\bibfnamefont {L.}~\bibnamefont {Chaix}},
  \bibinfo {author} {\bibfnamefont {E.~W.}\ \bibnamefont {Huang}}, \bibinfo
  {author} {\bibfnamefont {R.}~\bibnamefont {Fumagalli}}, \bibinfo {author}
  {\bibfnamefont {Y.~Y.}\ \bibnamefont {Peng}}, \bibinfo {author}
  {\bibfnamefont {B.}~\bibnamefont {Moritz}}, \bibinfo {author} {\bibfnamefont
  {K.}~\bibnamefont {Kummer}}, \bibinfo {author} {\bibfnamefont {N.~B.}\
  \bibnamefont {Brookes}}, \bibinfo {author} {\bibfnamefont {W.~C.}\
  \bibnamefont {Lee}}, \bibinfo {author} {\bibfnamefont {M.}~\bibnamefont
  {Hashimoto}}, \bibinfo {author} {\bibfnamefont {T.}~\bibnamefont {Sarkar}},
  \bibinfo {author} {\bibfnamefont {J.-F.}\ \bibnamefont {He}}, \bibinfo
  {author} {\bibfnamefont {C.~R.}\ \bibnamefont {Rotundu}}, \bibinfo {author}
  {\bibfnamefont {Y.~S.}\ \bibnamefont {Lee}}, \bibinfo {author} {\bibfnamefont
  {R.~L.}\ \bibnamefont {Greene}}, \bibinfo {author} {\bibfnamefont
  {L.}~\bibnamefont {Braicovich}}, \bibinfo {author} {\bibfnamefont
  {G.}~\bibnamefont {Ghiringhelli}}, \bibinfo {author} {\bibfnamefont {Z.~X.}\
  \bibnamefont {Shen}}, \bibinfo {author} {\bibfnamefont {T.~P.}\ \bibnamefont
  {Devereaux}},\ and\ \bibinfo {author} {\bibfnamefont {W.~S.}\ \bibnamefont
  {Lee}},\ }\href {https://doi.org/10.1038/s41586-018-0648-3} {\bibfield
  {journal} {\bibinfo  {journal} {Nature}\ }\textbf {\bibinfo {volume} {563}},\
  \bibinfo {pages} {374} (\bibinfo {year} {2018})}\BibitemShut {NoStop}%
\bibitem [{\citenamefont {Lin}\ \emph {et~al.}(2020)\citenamefont {Lin},
  \citenamefont {Yuan}, \citenamefont {Jin}, \citenamefont {Yin}, \citenamefont
  {Li}, \citenamefont {Zhou}, \citenamefont {Lu}, \citenamefont {Dantz},
  \citenamefont {Schmitt}, \citenamefont {Ding}, \citenamefont {Guo},
  \citenamefont {Dean},\ and\ \citenamefont {Liu}}]{Lin2020}%
  \BibitemOpen
  \bibfield  {author} {\bibinfo {author} {\bibfnamefont {J.}~\bibnamefont
  {Lin}}, \bibinfo {author} {\bibfnamefont {J.}~\bibnamefont {Yuan}}, \bibinfo
  {author} {\bibfnamefont {K.}~\bibnamefont {Jin}}, \bibinfo {author}
  {\bibfnamefont {Z.}~\bibnamefont {Yin}}, \bibinfo {author} {\bibfnamefont
  {G.}~\bibnamefont {Li}}, \bibinfo {author} {\bibfnamefont {K.-J.}\
  \bibnamefont {Zhou}}, \bibinfo {author} {\bibfnamefont {X.}~\bibnamefont
  {Lu}}, \bibinfo {author} {\bibfnamefont {M.}~\bibnamefont {Dantz}}, \bibinfo
  {author} {\bibfnamefont {T.}~\bibnamefont {Schmitt}}, \bibinfo {author}
  {\bibfnamefont {H.}~\bibnamefont {Ding}}, \bibinfo {author} {\bibfnamefont
  {H.}~\bibnamefont {Guo}}, \bibinfo {author} {\bibfnamefont {M.~P.~M.}\
  \bibnamefont {Dean}},\ and\ \bibinfo {author} {\bibfnamefont
  {X.}~\bibnamefont {Liu}},\ }\href {https://doi.org/10.1038/s41535-019-0205-9}
  {\bibfield  {journal} {\bibinfo  {journal} {npj Quantum Materials}\ }\textbf
  {\bibinfo {volume} {5}},\ \bibinfo {pages} {4} (\bibinfo {year}
  {2020})}\BibitemShut {NoStop}%
\bibitem [{\citenamefont {Lai}\ \emph {et~al.}(2014)\citenamefont {Lai},
  \citenamefont {Fung}, \citenamefont {Wu}, \citenamefont {Huang},
  \citenamefont {Fu}, \citenamefont {Lin}, \citenamefont {Huang}, \citenamefont
  {Chiu}, \citenamefont {Wang}, \citenamefont {Huang}, \citenamefont {Tseng},
  \citenamefont {Chung}, \citenamefont {Chen},\ and\ \citenamefont
  {Huang}}]{Lai2014}%
  \BibitemOpen
  \bibfield  {author} {\bibinfo {author} {\bibfnamefont {C.~H.}\ \bibnamefont
  {Lai}}, \bibinfo {author} {\bibfnamefont {H.~S.}\ \bibnamefont {Fung}},
  \bibinfo {author} {\bibfnamefont {W.~B.}\ \bibnamefont {Wu}}, \bibinfo
  {author} {\bibfnamefont {H.~Y.}\ \bibnamefont {Huang}}, \bibinfo {author}
  {\bibfnamefont {H.~W.}\ \bibnamefont {Fu}}, \bibinfo {author} {\bibfnamefont
  {S.~W.}\ \bibnamefont {Lin}}, \bibinfo {author} {\bibfnamefont {S.~W.}\
  \bibnamefont {Huang}}, \bibinfo {author} {\bibfnamefont {C.~C.}\ \bibnamefont
  {Chiu}}, \bibinfo {author} {\bibfnamefont {D.~J.}\ \bibnamefont {Wang}},
  \bibinfo {author} {\bibfnamefont {L.~J.}\ \bibnamefont {Huang}}, \bibinfo
  {author} {\bibfnamefont {T.~C.}\ \bibnamefont {Tseng}}, \bibinfo {author}
  {\bibfnamefont {S.~C.}\ \bibnamefont {Chung}}, \bibinfo {author}
  {\bibfnamefont {C.~T.}\ \bibnamefont {Chen}},\ and\ \bibinfo {author}
  {\bibfnamefont {D.~J.}\ \bibnamefont {Huang}},\ }\href
  {https://doi.org/10.1107/S1600577513030877} {\bibfield  {journal} {\bibinfo
  {journal} {J. Synchrotron Radiat.}\ }\textbf {\bibinfo {volume} {21}},\
  \bibinfo {pages} {325} (\bibinfo {year} {2014})}\BibitemShut {NoStop}%
\bibitem [{\citenamefont {Ament}\ \emph {et~al.}(2009)\citenamefont {Ament},
  \citenamefont {Ghiringhelli}, \citenamefont {Sala}, \citenamefont
  {Braicovich},\ and\ \citenamefont {van~den Brink}}]{Ament2009b}%
  \BibitemOpen
  \bibfield  {author} {\bibinfo {author} {\bibfnamefont {L.~J.~P.}\
  \bibnamefont {Ament}}, \bibinfo {author} {\bibfnamefont {G.}~\bibnamefont
  {Ghiringhelli}}, \bibinfo {author} {\bibfnamefont {M.~M.}\ \bibnamefont
  {Sala}}, \bibinfo {author} {\bibfnamefont {L.}~\bibnamefont {Braicovich}},\
  and\ \bibinfo {author} {\bibfnamefont {J.}~\bibnamefont {van~den Brink}},\
  }\href {https://doi.org/10.1103/PhysRevLett.103.117003} {\bibfield  {journal}
  {\bibinfo  {journal} {Phys. Rev. Lett.}\ }\textbf {\bibinfo {volume} {103}},\
  \bibinfo {eid} {117003} (\bibinfo {year} {2009})}\BibitemShut {NoStop}%
\bibitem [{\citenamefont {Sala}\ \emph {et~al.}(2011)\citenamefont {Sala},
  \citenamefont {Bisogni}, \citenamefont {Aruta}, \citenamefont {Balestrino},
  \citenamefont {Berger}, \citenamefont {Brookes}, \citenamefont {de~Luca},
  \citenamefont {Castro}, \citenamefont {Grioni}, \citenamefont {Guarise},
  \citenamefont {Medaglia}, \citenamefont {Granozio}, \citenamefont {Minola},
  \citenamefont {Perna}, \citenamefont {Radovic}, \citenamefont {Salluzzo},
  \citenamefont {Schmitt}, \citenamefont {Zhou}, \citenamefont {Braicovich},\
  and\ \citenamefont {Ghiringhelli}}]{Sala2011}%
  \BibitemOpen
  \bibfield  {author} {\bibinfo {author} {\bibfnamefont {M.~M.}\ \bibnamefont
  {Sala}}, \bibinfo {author} {\bibfnamefont {V.}~\bibnamefont {Bisogni}},
  \bibinfo {author} {\bibfnamefont {C.}~\bibnamefont {Aruta}}, \bibinfo
  {author} {\bibfnamefont {G.}~\bibnamefont {Balestrino}}, \bibinfo {author}
  {\bibfnamefont {H.}~\bibnamefont {Berger}}, \bibinfo {author} {\bibfnamefont
  {N.~B.}\ \bibnamefont {Brookes}}, \bibinfo {author} {\bibfnamefont {G.~M.}\
  \bibnamefont {de~Luca}}, \bibinfo {author} {\bibfnamefont {D.~D.}\
  \bibnamefont {Castro}}, \bibinfo {author} {\bibfnamefont {M.}~\bibnamefont
  {Grioni}}, \bibinfo {author} {\bibfnamefont {M.}~\bibnamefont {Guarise}},
  \bibinfo {author} {\bibfnamefont {P.~G.}\ \bibnamefont {Medaglia}}, \bibinfo
  {author} {\bibfnamefont {F.~M.}\ \bibnamefont {Granozio}}, \bibinfo {author}
  {\bibfnamefont {M.}~\bibnamefont {Minola}}, \bibinfo {author} {\bibfnamefont
  {P.}~\bibnamefont {Perna}}, \bibinfo {author} {\bibfnamefont
  {M.}~\bibnamefont {Radovic}}, \bibinfo {author} {\bibfnamefont
  {M.}~\bibnamefont {Salluzzo}}, \bibinfo {author} {\bibfnamefont
  {T.}~\bibnamefont {Schmitt}}, \bibinfo {author} {\bibfnamefont {K.~J.}\
  \bibnamefont {Zhou}}, \bibinfo {author} {\bibfnamefont {L.}~\bibnamefont
  {Braicovich}},\ and\ \bibinfo {author} {\bibfnamefont {G.}~\bibnamefont
  {Ghiringhelli}},\ }\href {http://stacks.iop.org/1367-2630/13/i=4/a=043026}
  {\bibfield  {journal} {\bibinfo  {journal} {New J. Phys.}\ }\textbf {\bibinfo
  {volume} {13}},\ \bibinfo {pages} {043026} (\bibinfo {year}
  {2011})}\BibitemShut {NoStop}%
\bibitem [{\citenamefont {Hill}\ \emph {et~al.}(1998)\citenamefont {Hill},
  \citenamefont {Kao}, \citenamefont {Caliebe}, \citenamefont {Matsubara},
  \citenamefont {Kotani}, \citenamefont {Peng},\ and\ \citenamefont
  {Greene}}]{Hill1998}%
  \BibitemOpen
  \bibfield  {author} {\bibinfo {author} {\bibfnamefont {J.~P.}\ \bibnamefont
  {Hill}}, \bibinfo {author} {\bibfnamefont {C.-C.}\ \bibnamefont {Kao}},
  \bibinfo {author} {\bibfnamefont {W.~A.~L.}\ \bibnamefont {Caliebe}},
  \bibinfo {author} {\bibfnamefont {M.}~\bibnamefont {Matsubara}}, \bibinfo
  {author} {\bibfnamefont {A.}~\bibnamefont {Kotani}}, \bibinfo {author}
  {\bibfnamefont {J.~L.}\ \bibnamefont {Peng}},\ and\ \bibinfo {author}
  {\bibfnamefont {R.~L.}\ \bibnamefont {Greene}},\ }\href
  {https://doi.org/10.1103/PhysRevLett.80.4967} {\bibfield  {journal} {\bibinfo
   {journal} {Phys. Rev. Lett.}\ }\textbf {\bibinfo {volume} {80}},\ \bibinfo
  {pages} {4967} (\bibinfo {year} {1998})}\BibitemShut {NoStop}%
\bibitem [{\citenamefont {Kim}\ \emph {et~al.}(2004)\citenamefont {Kim},
  \citenamefont {Hill}, \citenamefont {Gu}, \citenamefont {Chou}, \citenamefont
  {Wakimoto}, \citenamefont {Birgeneau}, \citenamefont {Komiya}, \citenamefont
  {Ando}, \citenamefont {Motoyama}, \citenamefont {Kojima}, \citenamefont
  {Uchida}, \citenamefont {Casa},\ and\ \citenamefont {Gog}}]{Kim2004b}%
  \BibitemOpen
  \bibfield  {author} {\bibinfo {author} {\bibfnamefont {Y.-J.}\ \bibnamefont
  {Kim}}, \bibinfo {author} {\bibfnamefont {J.~P.}\ \bibnamefont {Hill}},
  \bibinfo {author} {\bibfnamefont {G.~D.}\ \bibnamefont {Gu}}, \bibinfo
  {author} {\bibfnamefont {F.~C.}\ \bibnamefont {Chou}}, \bibinfo {author}
  {\bibfnamefont {S.}~\bibnamefont {Wakimoto}}, \bibinfo {author}
  {\bibfnamefont {R.~J.}\ \bibnamefont {Birgeneau}}, \bibinfo {author}
  {\bibfnamefont {S.}~\bibnamefont {Komiya}}, \bibinfo {author} {\bibfnamefont
  {Y.}~\bibnamefont {Ando}}, \bibinfo {author} {\bibfnamefont {N.}~\bibnamefont
  {Motoyama}}, \bibinfo {author} {\bibfnamefont {K.~M.}\ \bibnamefont
  {Kojima}}, \bibinfo {author} {\bibfnamefont {S.}~\bibnamefont {Uchida}},
  \bibinfo {author} {\bibfnamefont {D.}~\bibnamefont {Casa}},\ and\ \bibinfo
  {author} {\bibfnamefont {T.}~\bibnamefont {Gog}},\ }\href
  {https://doi.org/10.1103/PhysRevB.70.205128} {\bibfield  {journal} {\bibinfo
  {journal} {Phys. Rev. B}\ }\textbf {\bibinfo {volume} {70}},\ \bibinfo {eid}
  {205128} (\bibinfo {year} {2004})}\BibitemShut {NoStop}%
\bibitem [{\citenamefont {Uefuji}\ \emph {et~al.}(2002)\citenamefont {Uefuji},
  \citenamefont {Kurahashi}, \citenamefont {Fujita}, \citenamefont {Matsuda},\
  and\ \citenamefont {Yamada}}]{Uefuji2002}%
  \BibitemOpen
  \bibfield  {author} {\bibinfo {author} {\bibfnamefont {T.}~\bibnamefont
  {Uefuji}}, \bibinfo {author} {\bibfnamefont {K.}~\bibnamefont {Kurahashi}},
  \bibinfo {author} {\bibfnamefont {M.}~\bibnamefont {Fujita}}, \bibinfo
  {author} {\bibfnamefont {M.}~\bibnamefont {Matsuda}},\ and\ \bibinfo {author}
  {\bibfnamefont {K.}~\bibnamefont {Yamada}},\ }\href
  {https://doi.org/http://dx.doi.org/10.1016/S0921-4534(02)01426-0} {\bibfield
  {journal} {\bibinfo  {journal} {Physica C}\ }\textbf {\bibinfo {volume}
  {378-381}},\ \bibinfo {pages} {273 } (\bibinfo {year} {2002})}\BibitemShut
  {NoStop}%
\bibitem [{\citenamefont {Mang}\ \emph
  {et~al.}(2004{\natexlab{b}})\citenamefont {Mang}, \citenamefont {Vajk},
  \citenamefont {Arvanitaki}, \citenamefont {Lynn},\ and\ \citenamefont
  {Greven}}]{Mang2004a}%
  \BibitemOpen
  \bibfield  {author} {\bibinfo {author} {\bibfnamefont {P.~K.}\ \bibnamefont
  {Mang}}, \bibinfo {author} {\bibfnamefont {O.~P.}\ \bibnamefont {Vajk}},
  \bibinfo {author} {\bibfnamefont {A.}~\bibnamefont {Arvanitaki}}, \bibinfo
  {author} {\bibfnamefont {J.~W.}\ \bibnamefont {Lynn}},\ and\ \bibinfo
  {author} {\bibfnamefont {M.}~\bibnamefont {Greven}},\ }\href
  {https://doi.org/10.1103/PhysRevLett.93.027002} {\bibfield  {journal}
  {\bibinfo  {journal} {Phys. Rev. Lett.}\ }\textbf {\bibinfo {volume} {93}},\
  \bibinfo {pages} {027002} (\bibinfo {year} {2004}{\natexlab{b}})}\BibitemShut
  {NoStop}%
\bibitem [{\citenamefont {Kang}\ \emph {et~al.}(2019)\citenamefont {Kang},
  \citenamefont {Pelliciari}, \citenamefont {Krockenberger}, \citenamefont
  {Li}, \citenamefont {McNally}, \citenamefont {Paris}, \citenamefont {Liang},
  \citenamefont {Hardy}, \citenamefont {Bonn}, \citenamefont {Yamamoto},
  \citenamefont {Schmitt},\ and\ \citenamefont {Comin}}]{Kang2019}%
  \BibitemOpen
  \bibfield  {author} {\bibinfo {author} {\bibfnamefont {M.}~\bibnamefont
  {Kang}}, \bibinfo {author} {\bibfnamefont {J.}~\bibnamefont {Pelliciari}},
  \bibinfo {author} {\bibfnamefont {Y.}~\bibnamefont {Krockenberger}}, \bibinfo
  {author} {\bibfnamefont {J.}~\bibnamefont {Li}}, \bibinfo {author}
  {\bibfnamefont {D.~E.}\ \bibnamefont {McNally}}, \bibinfo {author}
  {\bibfnamefont {E.}~\bibnamefont {Paris}}, \bibinfo {author} {\bibfnamefont
  {R.}~\bibnamefont {Liang}}, \bibinfo {author} {\bibfnamefont {W.~N.}\
  \bibnamefont {Hardy}}, \bibinfo {author} {\bibfnamefont {D.~A.}\ \bibnamefont
  {Bonn}}, \bibinfo {author} {\bibfnamefont {H.}~\bibnamefont {Yamamoto}},
  \bibinfo {author} {\bibfnamefont {T.}~\bibnamefont {Schmitt}},\ and\ \bibinfo
  {author} {\bibfnamefont {R.}~\bibnamefont {Comin}},\ }\href
  {https://doi.org/10.1103/PhysRevB.99.045105} {\bibfield  {journal} {\bibinfo
  {journal} {Phys. Rev. B}\ }\textbf {\bibinfo {volume} {99}},\ \bibinfo
  {pages} {045105} (\bibinfo {year} {2019})}\BibitemShut {NoStop}%
\bibitem [{\citenamefont {Monney}\ \emph {et~al.}(2016)\citenamefont {Monney},
  \citenamefont {Schmitt}, \citenamefont {Matt}, \citenamefont {Mesot},
  \citenamefont {Strocov}, \citenamefont {Lipscombe}, \citenamefont {Hayden},\
  and\ \citenamefont {Chang}}]{Monney2016}%
  \BibitemOpen
  \bibfield  {author} {\bibinfo {author} {\bibfnamefont {C.}~\bibnamefont
  {Monney}}, \bibinfo {author} {\bibfnamefont {T.}~\bibnamefont {Schmitt}},
  \bibinfo {author} {\bibfnamefont {C.~E.}\ \bibnamefont {Matt}}, \bibinfo
  {author} {\bibfnamefont {J.}~\bibnamefont {Mesot}}, \bibinfo {author}
  {\bibfnamefont {V.~N.}\ \bibnamefont {Strocov}}, \bibinfo {author}
  {\bibfnamefont {O.~J.}\ \bibnamefont {Lipscombe}}, \bibinfo {author}
  {\bibfnamefont {S.~M.}\ \bibnamefont {Hayden}},\ and\ \bibinfo {author}
  {\bibfnamefont {J.}~\bibnamefont {Chang}},\ }\href
  {https://doi.org/10.1103/PhysRevB.93.075103} {\bibfield  {journal} {\bibinfo
  {journal} {Phys. Rev. B}\ }\textbf {\bibinfo {volume} {93}},\ \bibinfo
  {pages} {075103} (\bibinfo {year} {2016})}\BibitemShut {NoStop}%
\bibitem [{\citenamefont {Lu}\ \emph {et~al.}(2005)\citenamefont {Lu},
  \citenamefont {Chabot-Couture}, \citenamefont {Zhao}, \citenamefont
  {Hancock}, \citenamefont {Kaneko}, \citenamefont {Vajk}, \citenamefont {Yu},
  \citenamefont {Grenier}, \citenamefont {Kim}, \citenamefont {Casa},
  \citenamefont {Gog},\ and\ \citenamefont {Greven}}]{Lu2005}%
  \BibitemOpen
  \bibfield  {author} {\bibinfo {author} {\bibfnamefont {L.}~\bibnamefont
  {Lu}}, \bibinfo {author} {\bibfnamefont {G.}~\bibnamefont {Chabot-Couture}},
  \bibinfo {author} {\bibfnamefont {X.}~\bibnamefont {Zhao}}, \bibinfo {author}
  {\bibfnamefont {J.~N.}\ \bibnamefont {Hancock}}, \bibinfo {author}
  {\bibfnamefont {N.}~\bibnamefont {Kaneko}}, \bibinfo {author} {\bibfnamefont
  {O.~P.}\ \bibnamefont {Vajk}}, \bibinfo {author} {\bibfnamefont
  {G.}~\bibnamefont {Yu}}, \bibinfo {author} {\bibfnamefont {S.}~\bibnamefont
  {Grenier}}, \bibinfo {author} {\bibfnamefont {Y.~J.}\ \bibnamefont {Kim}},
  \bibinfo {author} {\bibfnamefont {D.}~\bibnamefont {Casa}}, \bibinfo {author}
  {\bibfnamefont {T.}~\bibnamefont {Gog}},\ and\ \bibinfo {author}
  {\bibfnamefont {M.}~\bibnamefont {Greven}},\ }\href
  {https://doi.org/10.1103/PhysRevLett.95.217003} {\bibfield  {journal}
  {\bibinfo  {journal} {Phys. Rev. Lett.}\ }\textbf {\bibinfo {volume} {95}},\
  \bibinfo {eid} {217003} (\bibinfo {year} {2005})}\BibitemShut {NoStop}%
\bibitem [{\citenamefont {Jia}\ \emph {et~al.}(2012)\citenamefont {Jia},
  \citenamefont {Chen}, \citenamefont {Sorini}, \citenamefont {Moritz},\ and\
  \citenamefont {Devereaux}}]{Jia2012}%
  \BibitemOpen
  \bibfield  {author} {\bibinfo {author} {\bibfnamefont {C.~J.}\ \bibnamefont
  {Jia}}, \bibinfo {author} {\bibfnamefont {C.-C.}\ \bibnamefont {Chen}},
  \bibinfo {author} {\bibfnamefont {A.~P.}\ \bibnamefont {Sorini}}, \bibinfo
  {author} {\bibfnamefont {B.}~\bibnamefont {Moritz}},\ and\ \bibinfo {author}
  {\bibfnamefont {T.~P.}\ \bibnamefont {Devereaux}},\ }\href
  {http://stacks.iop.org/1367-2630/14/i=11/a=113038} {\bibfield  {journal}
  {\bibinfo  {journal} {New J. Phys.}\ }\textbf {\bibinfo {volume} {14}},\
  \bibinfo {pages} {113038} (\bibinfo {year} {2012})}\BibitemShut {NoStop}%
\bibitem [{\citenamefont {Wakimoto}\ \emph {et~al.}(2013)\citenamefont
  {Wakimoto}, \citenamefont {Ishii}, \citenamefont {Kimura}, \citenamefont
  {Ikeuchi}, \citenamefont {Yoshida}, \citenamefont {Adachi}, \citenamefont
  {Casa}, \citenamefont {Fujita}, \citenamefont {Fukunaga}, \citenamefont
  {Gog}, \citenamefont {Koike}, \citenamefont {Mizuki},\ and\ \citenamefont
  {Yamada}}]{Wakimoto2013}%
  \BibitemOpen
  \bibfield  {author} {\bibinfo {author} {\bibfnamefont {S.}~\bibnamefont
  {Wakimoto}}, \bibinfo {author} {\bibfnamefont {K.}~\bibnamefont {Ishii}},
  \bibinfo {author} {\bibfnamefont {H.}~\bibnamefont {Kimura}}, \bibinfo
  {author} {\bibfnamefont {K.}~\bibnamefont {Ikeuchi}}, \bibinfo {author}
  {\bibfnamefont {M.}~\bibnamefont {Yoshida}}, \bibinfo {author} {\bibfnamefont
  {T.}~\bibnamefont {Adachi}}, \bibinfo {author} {\bibfnamefont
  {D.}~\bibnamefont {Casa}}, \bibinfo {author} {\bibfnamefont {M.}~\bibnamefont
  {Fujita}}, \bibinfo {author} {\bibfnamefont {Y.}~\bibnamefont {Fukunaga}},
  \bibinfo {author} {\bibfnamefont {T.}~\bibnamefont {Gog}}, \bibinfo {author}
  {\bibfnamefont {Y.}~\bibnamefont {Koike}}, \bibinfo {author} {\bibfnamefont
  {J.}~\bibnamefont {Mizuki}},\ and\ \bibinfo {author} {\bibfnamefont
  {K.}~\bibnamefont {Yamada}},\ }\href
  {https://doi.org/10.1103/PhysRevB.87.104511} {\bibfield  {journal} {\bibinfo
  {journal} {Phys. Rev. B}\ }\textbf {\bibinfo {volume} {87}},\ \bibinfo
  {pages} {104511} (\bibinfo {year} {2013})}\BibitemShut {NoStop}%
\bibitem [{\citenamefont {Dagan}\ and\ \citenamefont
  {Greene}(2007)}]{Dagan2007}%
  \BibitemOpen
  \bibfield  {author} {\bibinfo {author} {\bibfnamefont {Y.}~\bibnamefont
  {Dagan}}\ and\ \bibinfo {author} {\bibfnamefont {R.~L.}\ \bibnamefont
  {Greene}},\ }\href {https://doi.org/10.1103/PhysRevB.76.024506} {\bibfield
  {journal} {\bibinfo  {journal} {Phys. Rev. B}\ }\textbf {\bibinfo {volume}
  {76}},\ \bibinfo {pages} {024506} (\bibinfo {year} {2007})}\BibitemShut
  {NoStop}%
\bibitem [{\citenamefont {Li}\ \emph {et~al.}(2019)\citenamefont {Li},
  \citenamefont {Tabis}, \citenamefont {Tang}, \citenamefont {Yu},
  \citenamefont {Jaroszynski}, \citenamefont {Bari{\v s}i{\'c}},\ and\
  \citenamefont {Greven}}]{Li2019}%
  \BibitemOpen
  \bibfield  {author} {\bibinfo {author} {\bibfnamefont {Y.}~\bibnamefont
  {Li}}, \bibinfo {author} {\bibfnamefont {W.}~\bibnamefont {Tabis}}, \bibinfo
  {author} {\bibfnamefont {Y.}~\bibnamefont {Tang}}, \bibinfo {author}
  {\bibfnamefont {G.}~\bibnamefont {Yu}}, \bibinfo {author} {\bibfnamefont
  {J.}~\bibnamefont {Jaroszynski}}, \bibinfo {author} {\bibfnamefont
  {N.}~\bibnamefont {Bari{\v s}i{\'c}}},\ and\ \bibinfo {author} {\bibfnamefont
  {M.}~\bibnamefont {Greven}},\ }\href {https://doi.org/10.1126/sciadv.aap7349}
  {\bibfield  {journal} {\bibinfo  {journal} {Sci. Adv.}\ }\textbf {\bibinfo
  {volume} {5}},\ \bibinfo {pages} {eaap7349} (\bibinfo {year}
  {2019})}\BibitemShut {NoStop}%
\bibitem [{\citenamefont {Yu}\ \emph {et~al.}(2007)\citenamefont {Yu},
  \citenamefont {Liang}, \citenamefont {Li}, \citenamefont {Fujino},
  \citenamefont {Murakami}, \citenamefont {Takeuchi},\ and\ \citenamefont
  {Greene}}]{Yu2007}%
  \BibitemOpen
  \bibfield  {author} {\bibinfo {author} {\bibfnamefont {W.}~\bibnamefont
  {Yu}}, \bibinfo {author} {\bibfnamefont {B.}~\bibnamefont {Liang}}, \bibinfo
  {author} {\bibfnamefont {P.}~\bibnamefont {Li}}, \bibinfo {author}
  {\bibfnamefont {S.}~\bibnamefont {Fujino}}, \bibinfo {author} {\bibfnamefont
  {T.}~\bibnamefont {Murakami}}, \bibinfo {author} {\bibfnamefont
  {I.}~\bibnamefont {Takeuchi}},\ and\ \bibinfo {author} {\bibfnamefont
  {R.~L.}\ \bibnamefont {Greene}},\ }\href
  {https://doi.org/10.1103/PhysRevB.75.020503} {\bibfield  {journal} {\bibinfo
  {journal} {Phys. Rev. B}\ }\textbf {\bibinfo {volume} {75}},\ \bibinfo
  {pages} {020503} (\bibinfo {year} {2007})}\BibitemShut {NoStop}%
\bibitem [{\citenamefont {Horio}\ \emph
  {et~al.}(2018{\natexlab{a}})\citenamefont {Horio}, \citenamefont
  {Krockenberger}, \citenamefont {Yamamoto}, \citenamefont {Yokoyama},
  \citenamefont {Takubo}, \citenamefont {Hirata}, \citenamefont {Sakamoto},
  \citenamefont {Koshiishi}, \citenamefont {Yasui}, \citenamefont {Ikenaga},
  \citenamefont {Shin}, \citenamefont {Yamamoto}, \citenamefont {Wadati},\ and\
  \citenamefont {Fujimori}}]{Horio2018}%
  \BibitemOpen
  \bibfield  {author} {\bibinfo {author} {\bibfnamefont {M.}~\bibnamefont
  {Horio}}, \bibinfo {author} {\bibfnamefont {Y.}~\bibnamefont
  {Krockenberger}}, \bibinfo {author} {\bibfnamefont {K.}~\bibnamefont
  {Yamamoto}}, \bibinfo {author} {\bibfnamefont {Y.}~\bibnamefont {Yokoyama}},
  \bibinfo {author} {\bibfnamefont {K.}~\bibnamefont {Takubo}}, \bibinfo
  {author} {\bibfnamefont {Y.}~\bibnamefont {Hirata}}, \bibinfo {author}
  {\bibfnamefont {S.}~\bibnamefont {Sakamoto}}, \bibinfo {author}
  {\bibfnamefont {K.}~\bibnamefont {Koshiishi}}, \bibinfo {author}
  {\bibfnamefont {A.}~\bibnamefont {Yasui}}, \bibinfo {author} {\bibfnamefont
  {E.}~\bibnamefont {Ikenaga}}, \bibinfo {author} {\bibfnamefont
  {S.}~\bibnamefont {Shin}}, \bibinfo {author} {\bibfnamefont {H.}~\bibnamefont
  {Yamamoto}}, \bibinfo {author} {\bibfnamefont {H.}~\bibnamefont {Wadati}},\
  and\ \bibinfo {author} {\bibfnamefont {A.}~\bibnamefont {Fujimori}},\ }\href
  {https://doi.org/10.1103/PhysRevLett.120.257001} {\bibfield  {journal}
  {\bibinfo  {journal} {Phys. Rev. Lett.}\ }\textbf {\bibinfo {volume} {120}},\
  \bibinfo {pages} {257001} (\bibinfo {year} {2018}{\natexlab{a}})}\BibitemShut
  {NoStop}%
\bibitem [{\citenamefont {Horio}\ \emph
  {et~al.}(2018{\natexlab{b}})\citenamefont {Horio}, \citenamefont
  {Krockenberger}, \citenamefont {Koshiishi}, \citenamefont {Nakata},
  \citenamefont {Hagiwara}, \citenamefont {Kobayashi}, \citenamefont {Horiba},
  \citenamefont {Kumigashira}, \citenamefont {Irie}, \citenamefont {Yamamoto},\
  and\ \citenamefont {Fujimori}}]{Horio2018b}%
  \BibitemOpen
  \bibfield  {author} {\bibinfo {author} {\bibfnamefont {M.}~\bibnamefont
  {Horio}}, \bibinfo {author} {\bibfnamefont {Y.}~\bibnamefont
  {Krockenberger}}, \bibinfo {author} {\bibfnamefont {K.}~\bibnamefont
  {Koshiishi}}, \bibinfo {author} {\bibfnamefont {S.}~\bibnamefont {Nakata}},
  \bibinfo {author} {\bibfnamefont {K.}~\bibnamefont {Hagiwara}}, \bibinfo
  {author} {\bibfnamefont {M.}~\bibnamefont {Kobayashi}}, \bibinfo {author}
  {\bibfnamefont {K.}~\bibnamefont {Horiba}}, \bibinfo {author} {\bibfnamefont
  {H.}~\bibnamefont {Kumigashira}}, \bibinfo {author} {\bibfnamefont
  {H.}~\bibnamefont {Irie}}, \bibinfo {author} {\bibfnamefont {H.}~\bibnamefont
  {Yamamoto}},\ and\ \bibinfo {author} {\bibfnamefont {A.}~\bibnamefont
  {Fujimori}},\ }\href {https://doi.org/10.1103/PhysRevB.98.020505} {\bibfield
  {journal} {\bibinfo  {journal} {Phys. Rev. B}\ }\textbf {\bibinfo {volume}
  {98}},\ \bibinfo {pages} {020505} (\bibinfo {year}
  {2018}{\natexlab{b}})}\BibitemShut {NoStop}%
\bibitem [{\citenamefont {Asano}\ \emph {et~al.}(2018)\citenamefont {Asano},
  \citenamefont {Ishii}, \citenamefont {Matsumura}, \citenamefont {Tsuji},
  \citenamefont {Ina}, \citenamefont {Suzuki},\ and\ \citenamefont
  {Fujita}}]{Asano2018}%
  \BibitemOpen
  \bibfield  {author} {\bibinfo {author} {\bibfnamefont {S.}~\bibnamefont
  {Asano}}, \bibinfo {author} {\bibfnamefont {K.}~\bibnamefont {Ishii}},
  \bibinfo {author} {\bibfnamefont {D.}~\bibnamefont {Matsumura}}, \bibinfo
  {author} {\bibfnamefont {T.}~\bibnamefont {Tsuji}}, \bibinfo {author}
  {\bibfnamefont {T.}~\bibnamefont {Ina}}, \bibinfo {author} {\bibfnamefont
  {K.~M.}\ \bibnamefont {Suzuki}},\ and\ \bibinfo {author} {\bibfnamefont
  {M.}~\bibnamefont {Fujita}},\ }\href {https://doi.org/10.7566/JPSJ.87.094710}
  {\bibfield  {journal} {\bibinfo  {journal} {J. Phys. Soc. Jpn.}\ }\textbf
  {\bibinfo {volume} {87}},\ \bibinfo {pages} {094710} (\bibinfo {year}
  {2018})}\BibitemShut {NoStop}%
\bibitem [{\citenamefont {Lin}\ \emph {et~al.}(2019)\citenamefont {Lin},
  \citenamefont {Horio}, \citenamefont {Kawamata}, \citenamefont {Saito},
  \citenamefont {Koshiishi}, \citenamefont {Sakamoto}, \citenamefont {Zhang},
  \citenamefont {Yamamoto}, \citenamefont {Ikeda}, \citenamefont {Hirata},
  \citenamefont {Takubo}, \citenamefont {Wadati}, \citenamefont {Yasui},
  \citenamefont {Takagi}, \citenamefont {Ikenaga}, \citenamefont {Adachi},
  \citenamefont {Koike},\ and\ \citenamefont {Fujimori}}]{Lin2019}%
  \BibitemOpen
  \bibfield  {author} {\bibinfo {author} {\bibfnamefont {C.}~\bibnamefont
  {Lin}}, \bibinfo {author} {\bibfnamefont {M.}~\bibnamefont {Horio}}, \bibinfo
  {author} {\bibfnamefont {T.}~\bibnamefont {Kawamata}}, \bibinfo {author}
  {\bibfnamefont {S.}~\bibnamefont {Saito}}, \bibinfo {author} {\bibfnamefont
  {K.}~\bibnamefont {Koshiishi}}, \bibinfo {author} {\bibfnamefont
  {S.}~\bibnamefont {Sakamoto}}, \bibinfo {author} {\bibfnamefont
  {Y.}~\bibnamefont {Zhang}}, \bibinfo {author} {\bibfnamefont
  {K.}~\bibnamefont {Yamamoto}}, \bibinfo {author} {\bibfnamefont
  {K.}~\bibnamefont {Ikeda}}, \bibinfo {author} {\bibfnamefont
  {Y.}~\bibnamefont {Hirata}}, \bibinfo {author} {\bibfnamefont
  {K.}~\bibnamefont {Takubo}}, \bibinfo {author} {\bibfnamefont
  {H.}~\bibnamefont {Wadati}}, \bibinfo {author} {\bibfnamefont
  {A.}~\bibnamefont {Yasui}}, \bibinfo {author} {\bibfnamefont
  {Y.}~\bibnamefont {Takagi}}, \bibinfo {author} {\bibfnamefont
  {E.}~\bibnamefont {Ikenaga}}, \bibinfo {author} {\bibfnamefont
  {T.}~\bibnamefont {Adachi}}, \bibinfo {author} {\bibfnamefont
  {Y.}~\bibnamefont {Koike}},\ and\ \bibinfo {author} {\bibfnamefont
  {A.}~\bibnamefont {Fujimori}},\ }\href
  {https://doi.org/10.7566/JPSJ.88.115004} {\bibfield  {journal} {\bibinfo
  {journal} {J. Phys. Soc. Jpn.}\ }\textbf {\bibinfo {volume} {88}},\ \bibinfo
  {pages} {115004} (\bibinfo {year} {2019})}\BibitemShut {NoStop}%
\bibitem [{\citenamefont {Asano}\ \emph {et~al.}(2020)\citenamefont {Asano},
  \citenamefont {Ishii}, \citenamefont {Matsumura}, \citenamefont {Tsuji},
  \citenamefont {Kudo}, \citenamefont {Taniguchi}, \citenamefont {Saito},
  \citenamefont {Sunohara}, \citenamefont {Kawamata}, \citenamefont {Koike},\
  and\ \citenamefont {Fujita}}]{Asano2020}%
  \BibitemOpen
  \bibfield  {author} {\bibinfo {author} {\bibfnamefont {S.}~\bibnamefont
  {Asano}}, \bibinfo {author} {\bibfnamefont {K.}~\bibnamefont {Ishii}},
  \bibinfo {author} {\bibfnamefont {D.}~\bibnamefont {Matsumura}}, \bibinfo
  {author} {\bibfnamefont {T.}~\bibnamefont {Tsuji}}, \bibinfo {author}
  {\bibfnamefont {K.}~\bibnamefont {Kudo}}, \bibinfo {author} {\bibfnamefont
  {T.}~\bibnamefont {Taniguchi}}, \bibinfo {author} {\bibfnamefont
  {S.}~\bibnamefont {Saito}}, \bibinfo {author} {\bibfnamefont
  {T.}~\bibnamefont {Sunohara}}, \bibinfo {author} {\bibfnamefont
  {T.}~\bibnamefont {Kawamata}}, \bibinfo {author} {\bibfnamefont
  {Y.}~\bibnamefont {Koike}},\ and\ \bibinfo {author} {\bibfnamefont
  {M.}~\bibnamefont {Fujita}},\ }\Eprint {https://arxiv.org/abs/2005.10681v1}
  {arXiv:2005.10681v1 [cond-mat.supr-con]}  (\bibinfo {year}
  {2020})\BibitemShut {NoStop}%
\bibitem [{\citenamefont {Higgins}\ \emph {et~al.}(2006)\citenamefont
  {Higgins}, \citenamefont {Dagan}, \citenamefont {Barr}, \citenamefont
  {Weaver},\ and\ \citenamefont {Greene}}]{Higgins2006}%
  \BibitemOpen
  \bibfield  {author} {\bibinfo {author} {\bibfnamefont {J.~S.}\ \bibnamefont
  {Higgins}}, \bibinfo {author} {\bibfnamefont {Y.}~\bibnamefont {Dagan}},
  \bibinfo {author} {\bibfnamefont {M.~C.}\ \bibnamefont {Barr}}, \bibinfo
  {author} {\bibfnamefont {B.~D.}\ \bibnamefont {Weaver}},\ and\ \bibinfo
  {author} {\bibfnamefont {R.~L.}\ \bibnamefont {Greene}},\ }\href
  {https://doi.org/10.1103/PhysRevB.73.104510} {\bibfield  {journal} {\bibinfo
  {journal} {Phys. Rev. B}\ }\textbf {\bibinfo {volume} {73}},\ \bibinfo
  {pages} {104510} (\bibinfo {year} {2006})}\BibitemShut {NoStop}%
\bibitem [{\citenamefont {Arima}\ \emph {et~al.}(1993)\citenamefont {Arima},
  \citenamefont {Tokura},\ and\ \citenamefont {Uchida}}]{Arima1993}%
  \BibitemOpen
  \bibfield  {author} {\bibinfo {author} {\bibfnamefont {T.}~\bibnamefont
  {Arima}}, \bibinfo {author} {\bibfnamefont {Y.}~\bibnamefont {Tokura}},\ and\
  \bibinfo {author} {\bibfnamefont {S.}~\bibnamefont {Uchida}},\ }\href
  {https://doi.org/10.1103/PhysRevB.48.6597} {\bibfield  {journal} {\bibinfo
  {journal} {Phys. Rev. B}\ }\textbf {\bibinfo {volume} {48}},\ \bibinfo
  {pages} {6597} (\bibinfo {year} {1993})}\BibitemShut {NoStop}%
\bibitem [{\citenamefont {Greco}\ \emph {et~al.}(2016)\citenamefont {Greco},
  \citenamefont {Yamase},\ and\ \citenamefont {Bejas}}]{Greco2016}%
  \BibitemOpen
  \bibfield  {author} {\bibinfo {author} {\bibfnamefont {A.}~\bibnamefont
  {Greco}}, \bibinfo {author} {\bibfnamefont {H.}~\bibnamefont {Yamase}},\ and\
  \bibinfo {author} {\bibfnamefont {M.}~\bibnamefont {Bejas}},\ }\href
  {https://doi.org/10.1103/PhysRevB.94.075139} {\bibfield  {journal} {\bibinfo
  {journal} {Phys. Rev. B}\ }\textbf {\bibinfo {volume} {94}},\ \bibinfo
  {pages} {075139} (\bibinfo {year} {2016})}\BibitemShut {NoStop}%
\bibitem [{\citenamefont {Greco}\ \emph {et~al.}(2019)\citenamefont {Greco},
  \citenamefont {Yamase},\ and\ \citenamefont {Bejas}}]{Greco2019}%
  \BibitemOpen
  \bibfield  {author} {\bibinfo {author} {\bibfnamefont {A.}~\bibnamefont
  {Greco}}, \bibinfo {author} {\bibfnamefont {H.}~\bibnamefont {Yamase}},\ and\
  \bibinfo {author} {\bibfnamefont {M.}~\bibnamefont {Bejas}},\ }\href
  {https://doi.org/10.1038/s42005-018-0099-z} {\bibfield  {journal} {\bibinfo
  {journal} {Commun. Phys.}\ }\textbf {\bibinfo {volume} {2}},\ \bibinfo
  {pages} {3} (\bibinfo {year} {2019})}\BibitemShut {NoStop}%
\bibitem [{\citenamefont {Fujita}\ \emph {et~al.}(2003)\citenamefont {Fujita},
  \citenamefont {Kuroshima}, \citenamefont {Matsuda},\ and\ \citenamefont
  {Yamada}}]{Fujita2003}%
  \BibitemOpen
  \bibfield  {author} {\bibinfo {author} {\bibfnamefont {M.}~\bibnamefont
  {Fujita}}, \bibinfo {author} {\bibfnamefont {S.}~\bibnamefont {Kuroshima}},
  \bibinfo {author} {\bibfnamefont {M.}~\bibnamefont {Matsuda}},\ and\ \bibinfo
  {author} {\bibfnamefont {K.}~\bibnamefont {Yamada}},\ }\href
  {https://doi.org/DOI: 10.1016/S0921-4534(03)00750-0} {\bibfield  {journal}
  {\bibinfo  {journal} {Physica C}\ }\textbf {\bibinfo {volume} {392-396}},\
  \bibinfo {pages} {130 } (\bibinfo {year} {2003})}\BibitemShut {NoStop}%
\bibitem [{\citenamefont {Motoyama}\ \emph {et~al.}(2006)\citenamefont
  {Motoyama}, \citenamefont {Mang}, \citenamefont {Petitgrand}, \citenamefont
  {Yu}, \citenamefont {Vajk}, \citenamefont {Vishik},\ and\ \citenamefont
  {Greven}}]{Motoyama2006}%
  \BibitemOpen
  \bibfield  {author} {\bibinfo {author} {\bibfnamefont {E.~M.}\ \bibnamefont
  {Motoyama}}, \bibinfo {author} {\bibfnamefont {P.~K.}\ \bibnamefont {Mang}},
  \bibinfo {author} {\bibfnamefont {D.}~\bibnamefont {Petitgrand}}, \bibinfo
  {author} {\bibfnamefont {G.}~\bibnamefont {Yu}}, \bibinfo {author}
  {\bibfnamefont {O.~P.}\ \bibnamefont {Vajk}}, \bibinfo {author}
  {\bibfnamefont {I.~M.}\ \bibnamefont {Vishik}},\ and\ \bibinfo {author}
  {\bibfnamefont {M.}~\bibnamefont {Greven}},\ }\href
  {https://doi.org/10.1103/PhysRevLett.96.137002} {\bibfield  {journal}
  {\bibinfo  {journal} {Phys. Rev. Lett.}\ }\textbf {\bibinfo {volume} {96}},\
  \bibinfo {pages} {137002} (\bibinfo {year} {2006})}\BibitemShut {NoStop}%
\bibitem [{\citenamefont {Yu}\ \emph {et~al.}(2010)\citenamefont {Yu},
  \citenamefont {Li}, \citenamefont {Motoyama}, \citenamefont {Hradil},
  \citenamefont {Mole},\ and\ \citenamefont {Greven}}]{Yu2010}%
  \BibitemOpen
  \bibfield  {author} {\bibinfo {author} {\bibfnamefont {G.}~\bibnamefont
  {Yu}}, \bibinfo {author} {\bibfnamefont {Y.}~\bibnamefont {Li}}, \bibinfo
  {author} {\bibfnamefont {E.~M.}\ \bibnamefont {Motoyama}}, \bibinfo {author}
  {\bibfnamefont {K.}~\bibnamefont {Hradil}}, \bibinfo {author} {\bibfnamefont
  {R.~A.}\ \bibnamefont {Mole}},\ and\ \bibinfo {author} {\bibfnamefont
  {M.}~\bibnamefont {Greven}},\ }\href
  {https://doi.org/10.1103/PhysRevB.82.172505} {\bibfield  {journal} {\bibinfo
  {journal} {Phys. Rev. B}\ }\textbf {\bibinfo {volume} {82}},\ \bibinfo
  {pages} {172505} (\bibinfo {year} {2010})}\BibitemShut {NoStop}%
\bibitem [{\citenamefont {Qazilbash}\ \emph {et~al.}(2005)\citenamefont
  {Qazilbash}, \citenamefont {Koitzsch}, \citenamefont {Dennis}, \citenamefont
  {Gozar}, \citenamefont {Balci}, \citenamefont {Kendziora}, \citenamefont
  {Greene},\ and\ \citenamefont {Blumberg}}]{Qazilbash2005}%
  \BibitemOpen
  \bibfield  {author} {\bibinfo {author} {\bibfnamefont {M.~M.}\ \bibnamefont
  {Qazilbash}}, \bibinfo {author} {\bibfnamefont {A.}~\bibnamefont {Koitzsch}},
  \bibinfo {author} {\bibfnamefont {B.~S.}\ \bibnamefont {Dennis}}, \bibinfo
  {author} {\bibfnamefont {A.}~\bibnamefont {Gozar}}, \bibinfo {author}
  {\bibfnamefont {H.}~\bibnamefont {Balci}}, \bibinfo {author} {\bibfnamefont
  {C.~A.}\ \bibnamefont {Kendziora}}, \bibinfo {author} {\bibfnamefont {R.~L.}\
  \bibnamefont {Greene}},\ and\ \bibinfo {author} {\bibfnamefont
  {G.}~\bibnamefont {Blumberg}},\ }\href
  {https://doi.org/10.1103/PhysRevB.72.214510} {\bibfield  {journal} {\bibinfo
  {journal} {Phys. Rev. B}\ }\textbf {\bibinfo {volume} {72}},\ \bibinfo
  {pages} {214510} (\bibinfo {year} {2005})}\BibitemShut {NoStop}%
\end{thebibliography}%

\end{document}